\documentclass[10pt,final]{IEEEtran}
\usepackage{amsfonts}
\usepackage{mathrsfs}
\usepackage{amssymb}
\usepackage{graphicx}
\usepackage{psfrag}
\usepackage{amsmath}
\usepackage{array}
\usepackage{cases}
\usepackage{cite}
\usepackage{tabularx}
\usepackage{url}
\usepackage{color}
\usepackage{soul}
\usepackage[level=0]{wgroup_message}
\makeatletter
\def\hlinew#1{%
  \noalign{\ifnum0=`}\fi\hrule \@height #1 \futurelet
   \reserved@a\@xhline}
\makeatother

\newcommand{\mnc}[3]{{#1}^{(\!\!\;#2\!\!\;)}_{#3}}
\newcommand{\MNC}[3]{{\mathbf{#1}}^{(\!\!\;#2\!\!\;)}_{#3}}

\newcommand{\hgap}{\hspace{2mm}}
\newcommand{\ml}{\hspace{-.02mm}\ell\hspace{0.3mm}}

\makeatletter
\renewcommand*{\@opargbegintheorem}[3]{\@IEEEtmpitemindent\itemindent\topsep 0pt\rmfamily \trivlist%
      \item[\hskip \labelsep{\indent\itshape #1\ #2}] \textit{(#3):}\ \itemindent\@IEEEtmpitemindent}
\makeatother
\newcommand{\paperTitle}{Generalized Interference Alignment --- Part I: Theoretical Framework}

\usepackage{varioref}

\begin{document}

\author{Liangzhong~Ruan,~\IEEEmembership{Member,~IEEE},
Vincent~K.N.~Lau, \IEEEmembership{Fellow,~IEEE},
and Moe~Z.~Win, \IEEEmembership{Fellow,~IEEE}

\thanks{L.~Ruan and M.~Z.~Win are with the Laboratory for Information and Decision Systems
(LIDS), MIT (e-mail: \texttt{lruan, moewin@mit.edu)}.}
\thanks{V.~K.~N.~Lau is with the ECE Department, HKUST (e-mail: \texttt{eeknlau@ust.hk}).}
}

\title{\paperTitle}
\newtheorem{Thm}{Theorem}[section]
\newtheorem{Lem}{Lemma}[section]
\newtheorem{Asm}{Assumption}[section]
\newtheorem{Def}{Definition}
\newtheorem{Remark}{\\ Remark}[section]
\newtheorem{Prob}{Problem}[section]
\newtheorem{Prop}{Proposition}[section]
\newtheorem{Alg}{Algorithm}
\newtheorem{Test}{Test}
\newtheorem{Cor}{Corollary}[section]
\definecolor{mblue}{rgb}{0.05,0.05,0.6}

\maketitle
\begin{abstract}
Interference alignment  (IA) has attracted enormous research interest as it achieves optimal capacity scaling with respect to signal to noise ratio on interference networks.
IA has also recently emerged as an effective tool in engineering interference
for secrecy protection on wireless wiretap networks.
However, despite the numerous works dedicated to IA, two of its fundamental
issues, i.e., feasibility conditions and transceiver design, are not completely
addressed in the literature.
In this two part paper, a generalized
interference alignment (GIA) technique is proposed to enhance the IA's capability in secrecy protection.
A theoretical framework is established to analyze the two fundamental
issues of GIA in Part I and then
 the performance of GIA in large-scale stochastic networks is characterized to
illustrate how GIA benefits secrecy protection in Part II.
The theoretical framework for GIA adopts methodologies from algebraic geometry,
determines the necessary and sufficient feasibility conditions of GIA, and generates a set of algorithms that  can solve the GIA problem.
This framework sets
up a foundation for the development and implementation of GIA.
\end{abstract}
\renewcommand{\IEEEQED}{\IEEEQEDopen}

\section{Introduction}
\mysubnote{The pros and cons of IA}
\subsection{Background and Survey}

Due to the broadcast nature of the wireless propagation medium, interference is a major factor that limits the performance of wireless communication networks.
Conventional interference control schemes, most of which adopt the principle of channel orthogonalization are in general non-capacity achieving  \cite{Sas:04,EtkTseWan:08}.
IA \cite{MadMotKha:08} reduces the effect of aggregated interference by aligning interference from multiple sources into lower-dimensional subspaces at receivers.
It achieves the optimal capacity scaling with respect to signal to noise ratio (SNR) in a wide range of networks \cite{CadJaf:J08,CadJafSha:09,CadJaf:09}.
On the other hand, in a wireless network that requires the secure exchange of confidential messages,  interference, which  enables legitimate partners to impede the eavesdropping receivers (ERs),  emerges as a potentially valuable resource for wireless network secrecy \cite{RabConWin:C12,LeeShiWin:C12}.
In order to impede the ERs without interfering with legitimate receivers (LRs), a few studies have adopted the IA scheme proposed in \cite{CadJaf:J08} to promote wireless secrecy \cite{KoyGamLaiPoo:11,BasUlu:12,KoyKokElG:12}.
However, the scheme in \cite{CadJaf:J08} is based on infinite-dimensional symbol extension, making it difficult to implement in practice.

\mysubnote{IA without symbol extension is practical. However, its theoretical foundation is incomplete.}
To avoid the infinite-dimension issue, researchers have developed spatial-domain IA schemes, in which interference is coordinated and canceled via the finite signal dimension provided by multiple antennas.
For this scheme, there are two fundamental issues:
(1) When is IA (without symbol extension) feasible;
and (2) Given that IA is feasible, how to design an algorithm to find transceivers that cancel all interference?
For the feasibility issue, the pioneering works characterize the IA feasibility conditions under some special configurations \cite{YetGouJaf:10,BreCarTse:11,RazGenLuo:12,WanGouJaf:12a,WanGouJaf:12b}.
In \cite{GonSanBel:12,GonCarSan:14}, a numerical test that checks IA feasibility is proposed.
In the authors' prior work \cite{RuaLauWin:J13}, we prove a sufficient IA feasibility condition for MIMO interference networks with a general configuration.
This results unifies and extends those in \cite{YetGouJaf:10,BreCarTse:11,RazGenLuo:12}.
For the transceiver design issue, there are two categories of algorithms: constructive ones and iterative ones. The constructive algorithms apply to networks with special configurations \cite{LeeParKim:09,TreGuiRie:09,KhaCho:12}.
The iterative algorithms \cite{GomCadJaf:08,PetHea:09,PetHea:11,GomCadJaf:11,SchShiBerHonUts:09,KumXue:10,GonSanHeaPet:10} apply to networks with general configurations, but they converge to local optimums.
Table~\ref{tab:feasibility} and~\ref{tab:algorithm} in Section~\ref{sec:model} summarize the contributions and limitations of the existing works on IA feasibility analysis and transceiver design.
The incomplete theoretical foundation imposes a great challenge on the development of IA.

\mysubnote{The value of interference in secrecy protection}
Furthermore, as will be discussed in detail in Part II, to promote the capability of IA in secrecy protection, it is desirable to introduce legitimate jammers (LJs) into the network and jointly coordinate the transmission policy of all legitimate partners to create stronger interference at the ERs without affecting the LRs.
In this paper, this technique is referred to as GIA.
To develop such a technique, the following challenges need to be addressed:

\begin{itemize}
\item{\bf Determine the feasibility conditions of GIA:}
Feasibility analysis of IA is challenging because
IA constraints are sets of non-linear equations, for which no systematic tool exists to analyze the feasible region.
In the authors' prior work \cite{RuaLauWin:J13}, by exploiting the connection between the feasibility of IA and the linear independence of the first order terms of IA constraints, an algebraic framework was established which gives a sufficient condition of IA feasibility.
However, this framework is incomplete as it does not characterize necessary feasibility conditions.
\item{\bf Design GIA transceivers under general configuration:}
For networks with a general configuration,
existing IA transceiver design algorithms
may not be able to find a solution even when IA is feasible.
The IA transceiver design problem is usually formulated into an interference leakage minimization form
\cite{GomCadJaf:08,PetHea:09} or a rank minimization form \cite{PapDim:12}. However, in both forms, the problem is non-convex, making it challenging to find solutions.
Moreover, in a network with many nodes, the dimension of the transceiver matrices is large.
Designing algorithms to solve a non-convex, high-dimensional problem is difficult.
\end{itemize}

\subsection{Contribution of This Work}
In this work, we will address the challenges listed above.
We will consider MIMO wireless-tap networks\footnote{ ``wireless wiretap" is referred to as ``wireless-tap" in this paper to emphasize
the wireless nature of the propagation medium.} with LJs.
By adopting tools from algebraic geometry \cite{Sha:96}, we establish a framework which shows the (almost sure) equivalence of the feasibility of the GIA transceiver design problem, the algebraic independence of GIA constraints, and the linear independence of the first order terms of GIA constraints.
This framework enables us to propose and prove a necessary and sufficient condition for GIA to be feasible in MIMO networks with a general configuration.
By combining this condition with graph theory \cite{HsuLin:B09}, we generate several insights into the relation between network configuration and GIA feasibility.

To address the challenge in GIA transceiver design, we exploit the equivalence between algebraic independence of GIA constraints and full rankness of their Jacobian matrix, and prove that when GIA is feasible, in a set of corresponding interference minimization problems, there is no performance gap between local and global optimums.
This fact enables us to find solutions for the GIA transceiver design problem by adopting existing local search algorithms.
The feasibility analysis and transceiver  design for GIA covers those for IA as a special case.

\subsection{Organization}
Section~\ref{sec:model} formulates the GIA problem.
Section~\ref{sec:background} introduces the mathematical preliminaries.
Section~\ref{sec:result} establishes an algebraic framework that determines GIA feasibility conditions and design GIA algorithms.
Section~\ref{sec:sim} provides numerical tests on the convergence issue of the proposed GIA transceiver design algorithm.
Finally, Section~\ref{sec:conclude} gives the conclusion.

\mynote{Notation}
\subsection{Notations}
\subsubsection{General}
$a$, $\mathbf{a}$, $\mathbf{A}$, and $\mathcal{A}$ represent scalar, vector, matrix, and set/space, respectively. $\mathbb{N}$, $\mathbb{Z}$, $\mathbb{R}$ and $\mathbb{C}$ denote the set of natural numbers, integers, real numbers, and complex numbers, respectively.

\subsubsection{Functions} $n|m$ denotes that $n$ divides $m$, and $n\;\mathrm{mod}\;m$ denotes $n$ modulo $m$, $n,m\in\mathbb{Z}$.
$\mathbb{I}\{\cdot\}$ is the indicator function.
$\begin{pmatrix}n\vspace{-2mm}\\m\end{pmatrix}$ is the Binomial coefficient with parameters $n,m\in\mathbb{N}$. $|a|$ represents the absolute value of scalar $a$, and $|\mathcal{A}|$ represents the cardinality of set $\mathcal{A}$.

\subsubsection{Linear algebra}
The operators $(\cdot)^\mathrm{T}$, $(\cdot)^\mathrm{H}$, $\det(\cdot)$, $\mathrm{rank}(\cdot)$, $||\cdot||_{\mathrm{F}}$, $\mathrm{trace}(\cdot)$, $(\cdot)^\sharp$, $\mathcal{N}(\cdot)$, and $\mathrm{vec}(\cdot)$ denote transpose, Hermitian transpose, determinant, rank, Frobenius norm, trace, Moore--Penrose pseudo inverse, null space, and  vectorization of a matrix.
$\mathrm{span}(\mathbf{A})$ and $\mathrm{span}(\{\mathbf{a}\})$ denote the linear space spanned by the column vectors of $\mathbf{A}$ and the vectors in set $\{\mathbf{a}\}$, respectively.
$\dim(\cdot)$ denotes the dimension of a space.
$\mathrm{diag}^{n}(\mathbf{A},\ldots,\mathbf{X})$ represents a block diagonal matrix with submatrices $\mathbf{A},\ldots,\mathbf{X}$ on its  $n$-th diagonal. For instance, $\mathrm{diag}^{-\!1}([2,1],[1,2])=\Bigg[\!\!{\scriptsize\begin{array}{*{5}{c@{\,}}c}0&0&0&0&0&0\\2&1&0&0&0&0\\0&0&1&2&0&0\end{array}}\!\!\Bigg]$. $\mathrm{diag}(\mathbf{A},\ldots,\mathbf{X})=\mathrm{diag}^{0}(\mathbf{A},\ldots,\mathbf{X})$, and $\mathrm{diag}[m](\mathbf{A})=\mathrm{diag}(\underbrace{\mathbf{A},\ldots,\mathbf{A}}_{m\mbox{ \scriptsize times}})$.

\subsubsection{Algebraic geometry}
For a field $\mathcal{K}$, $\mathcal{K}(x_1,\ldots,x_j)$ represents the field of rational functions in variables $x_1,\ldots,x_j$ with coefficients drawn from $\mathcal{K}$.
Notation $\langle f_1,\ldots,f_L\rangle$ denotes the ideal generated by polynomials $f_1,\ldots,f_S$; notation $\mathcal{V}(\cdot)$ denotes vanishing set of an ideal; and  notation $\mathbf{J}_{\mathbf{x}}(f_1,\ldots,f_L)$ represents the Jacobian matrix of polynomials $f_1,\ldots,f_L\in \mathcal{K}(x_1,\ldots,x_S)$ evaluated at point $\mathbf{x}\in\mathcal{K}^S$.

\section{Problem Formulation}
\label{sec:model}
In this section, the system model of wireless-tap networks is described, which is a generalization of interference networks, and then the GIA transceiver design problem is formulated.
\subsection{System Model}
\label{subsec:model}
\begin{figure}[t] \centering
\includegraphics[scale=0.75]{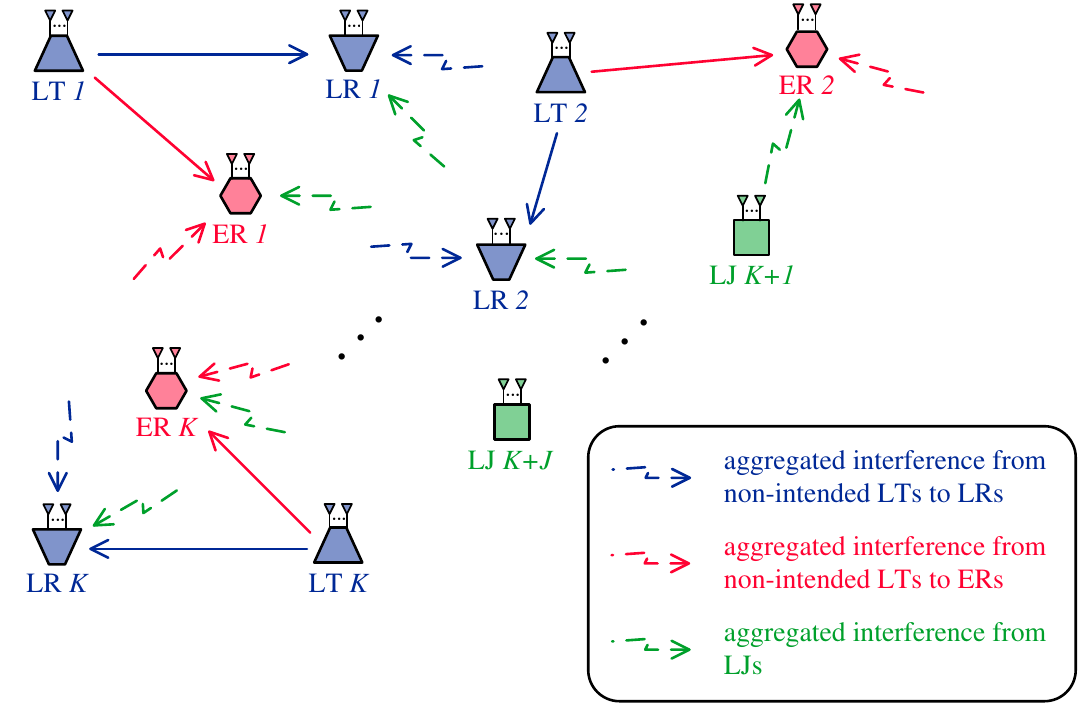}
\caption{Network configuration of wireless-tap networks with LJs.}
\label{fig_ChannelModel}
\end{figure}

Consider a network consisting of $K$ legitimate transmitter (LT)-LR pairs, $J$ LJs and $K$ ERs,\footnote{In
fact, as the proposed GIA technique does not require the channel state of the eavesdropping network, the ERs are not involved in GIA feasibility analysis. However, they remain in the system model to make the
notation consistent with Part II.}  as illustrated in Fig.~\ref{fig_ChannelModel}. (Note that LTs and LJs are indexed from 1 to $K$ and from $K\!+\!1$ to $K\!+\!J$, respectively.) Suppose LT $j$ (or LJ $j$, if $j>K$), LR $k$, and ER $k$ are equipped with $M_j$, $\mnc{N}{\ml}{k}$, and $\mnc{N}{e}{k}$ antennas, respectively. At each time slot, LT (or LJ) $j$ sends $d_j$ independent symbols.
LT $k$ attempts to send confidential messages to LR $k$, while ER $k$ attempts to intercept these messages. LJ~$j$ transmits dummy data to generate interference.

The received signals $\MNC{y}{\ml}{k},\MNC{y}{e}{k}\in \mathbb{C}^{d_k}$ at LR $k$ and ER $k$ are given by
\begin{eqnarray}
\MNC{y}{\iota}{k}=(\MNC{U}{\iota}{k})^\mathrm{H}\bigg(\mathbf{H}^{{(\!\!\;\iota\!\!\;)}}_{kk}
\mathbf{V}_{k}\mathbf{x}_{k} + \sum_{j=1,\neq k}^{\tilde{K}}
\mathbf{H}^{{(\!\!\;\iota\!\!\;)}}_{kj} \mathbf{V}_{j}\mathbf{x}_{j}
+\mathbf{z}^{{(\!\!\;\iota\!\!\;)}}_k\bigg),
\label{eqn:LRsignal}
\end{eqnarray}
where $\tilde{K}=K+J$, $\MNC{H}{\iota}{kj}\in \mathbb{C}^{N^{(\!\!\;\iota\!\!\;)}_k\times M_j}$, ${\iota}\in\{\ell,{e}\}$ are the channel matrices from LT (or LJ) $j$ to LR $k$ or ER $k$, whose entries are independent random variables drawn from continuous distributions; $\mathbf{x}_j \in \mathbb{C}^{d_j}$  is the encoded
information symbol at LT (or LJ) $j$; $\mathbf{V}_{j}\in \mathbb{C}^{M_j\times d_j}$ is the precoder at LT (or LJ) $j$; $\mathbf{U}^{(\!\!\;\iota\!\!\;)}_{k}\in \mathbb{C}^{N^{(\!\!\;\iota\!\!\;)}_k\times d_k}$, ${\iota}\in\{\ell,{e}\}$ is the decoder at LR
$k$ or ER $k$; and $\MNC{z}{\iota}{k}\in \mathbb{C}^{\mnc{N}{\iota}{k}\times 1}$, ${\iota}\in\{\ell,{e}\}$ is the
white Gaussian noise with zero mean and unit variance. The transmission power of LT (or LJ) $j$ is given by
\begin{eqnarray}
P_j = \mathbb{E}\left\{\mathrm{trace}\big(\mathbf{x}^\mathrm{H}_j\mathbf{V}^\mathrm{H}_j\mathbf{V}_j\mathbf{x}_j\big) \right\}.\label{eqn:power}
\end{eqnarray}
Define the configuration of the legitimate network as $\chi\triangleq\{(M_1,M_2,\ldots,M_{\tilde{K}}),(\mnc{N}{\ml}{1},\mnc{N}{\ml}{2},\ldots,
\mnc{N}{\ml}{K}),$ $(d_1,d_2,\ldots,d_{\tilde{K}})\}$.

\begin{Remark}[Applicability to Interference Networks] The wireless-tap network proposed above is a generalization of interference networks.
Specifically, when there is no LJ, i.e., $\tilde{K}=K$, and the channel state of the eavesdropping links are zero matrices, i.e., $\MNC{H}{\ml}{kj}=\mathbf{0}$, $\forall k,j$, the channel model \eqref{eqn:LRsignal} is reduced to that of conventional MIMO interference networks.
Hence, as further illustrated in Remark~\ref{remark:IA}, the theoretical results obtained in this work apply to MIMO interference networks.
~\hfill~\IEEEQED
\end{Remark}

\subsection{GIA Transceiver Design with Flexible Alignment Set}
Classical IA requires canceling interference on all cross links.
However, in large-scale networks this target may be infeasible and unnecessary.
On one hand, the limited policy space in transceiver design may be insufficient to cancel interference on all cross links; on the other hand, some links may have very deep fading and hence there is no need to cancel interference on these links.
Hence, to develop GIA strategies that fit large-scale networks, a more flexible approach must be adopted, in which the legitimate partners selectively cancel interference on a subset of cross links.
This problem is formulated as follows:
\begin{Prob}[GIA Transceiver Design]\label{pro:GIA}
Design transceivers $\{\MNC{U}{\ml}{k},$ $ \mathbf{V}_j\}$, $k\in\{1,\ldots,K\}$, $j\in\{1,\ldots,\tilde{K}\}$ that satisfy the following constraints:
\begin{eqnarray}
\hspace{-4mm}\mathrm{rank}\left((\MNC{U}{\ml}{k})^\mathrm{H}\MNC{H}{\ml}{kk}\mathbf{V}_k\right) \!\!&=\!\!& d_k\label{eqn:drank}, \quad\forall k\in\{1,\ldots,K\},\\
\mathrm{rank}\left(\mathbf{V}_j\right) \!\!&=\!\!& d_j\label{eqn:rankJ}, \quad\forall j\in\{K\hspace{-0.7mm}+\!1,\ldots,\tilde{K}\},\\
\hspace{-4mm}\mbox{and}\hspace{3.5mm}\quad(\MNC{U}{\ml}{k})^\mathrm{H}\MNC{H}{\ml}{kj}\mathbf{V}_j \!\!&=\!\!& \mathbf{0},\quad\hspace{1.2mm}\forall (k,j)\in \mathcal{A},\label{eqn:czero}
\end{eqnarray}
where $\mathcal{A}\subseteq\mathcal{A}_{\mathrm{all}}=\{(k,j):k\in\{1,\ldots,K\},j\in\{1,\ldots,\tilde{K}\},k\neq j\}$ is the alignment set.
It characterizes the set of cross links on which interference is to be canceled.~\hfill~\IEEEQED
\end{Prob}

\begin{Remark}[Connection between IA and GIA Problems]\label{remark:IA}
When there are no LJs and the alignment set includes all cross links, i.e., $\tilde{K}=K$, $\mathcal{A}=\mathcal{A}_{\mathrm{all}}$, Problem~\ref{pro:GIA} is converted to the classical IA problem on MIMO interference networks \cite{YetGouJaf:10} (without symbol extension).
Since Problem~\ref{pro:GIA} is a generalization of the classical IA problem, the feasibility conditions and algorithm design that  are obtained in Section~\ref{sec:result} naturally apply to the IA problem.
~\hfill~\IEEEQED
\end{Remark}

\begin{table}[h]
\caption{Applicable Configurations of Existing Necessary and Sufficient IA Feasibility Conditions}\label{tab:feasibility}
\centerline{
\begin{tabular}{c@{\hspace{-1.3mm}}c|l@{\quad}l@{\quad}l@{\quad}lc@{\hspace{-1.3mm}}}
\hlinew{0.8pt}
&Reference& \multicolumn{3}{c}{Network Configuration}&\\
\cline{2-5}
&\cite{YetGouJaf:10}&$K\in\mathbb{N},$&$d_k=1$,&$\forall k$&\\
\cline{2-5}
&\cite{BreCarTse:11}&$K\ge 3$,&$d_k=d$,&$\mnc{N}{\ml}{k}=M_k=N$,$\quad\forall k$& \\
\cline{2-5}
&\cite{RazGenLuo:12}&$K\in\mathbb{N}$,&$d_k=d$,&$d|\mnc{N}{\ml}{k}$, and $d|M_k$,$\quad\forall k$&\\
\cline{2-5}
&\cite{WanGouJaf:12a}&$K=3$,&$d_k=d$,&$\mnc{N}{\ml}{k}=N$, $M_k=M$,$\quad\forall k$&\\
\cline{2-5}
&&$K\in\mathbb{N}$,&$d_k=d$,&$\mnc{N}{\ml}{k}=N$, $M_k=M$,& \\
&&\multicolumn{2}{l}{$\min\{M,N\}\ge 2d$,} &$\forall k$ (extension of \cite{BreCarTse:11})&\\
&\raisebox{3mm}[0pt]{\hspace{-0.01mm}\cite{RuaLauWin:J13}}&$K\in\mathbb{N}$;&$d_k=d$,&$d|\mnc{N}{\ml}{k}$, or $d|M_k$,$\quad\forall k$&\\
&&\multicolumn{3}{l} {(extension of \cite{YetGouJaf:10,RazGenLuo:12})}&\\
\hlinew{0.8pt}
\end{tabular}
}
\end{table}
\begin{table}[h]
\caption{Applicable Configurations of Existing IA Algorithms}\label{tab:algorithm}
\centerline{
\begin{tabular}{c@{\hspace{-1.3mm}}c|c|l@{\;\;}l@{\;\;}lc@{\hspace{-1.3mm}}}
\hlinew{0.8pt}
&Reference&Type& \multicolumn{3}{c}{Network Configuration}&\\
\cline{2-6}
&\cite{CadJaf:J08} &constructive &$K=3$,&$d_k=d$, &$\mnc{N}{\ml}{k}=M_k=N,\;\;\forall k$&\\
\cline{2-6}
&\cite{WanGouJaf:12a}&constructive&$K=3$,&$d_k=d$, &$\mnc{N}{\ml}{k}=N$, $M_k=M,\;\;\forall k$&\\
\cline{2-6}
&\cite{LeeParKim:09} &constructive&$K\in\mathbb{N}$,&$d_k=1$, &$\mnc{N}{\ml}{k}=2,\;\;\forall k$&\\
\cline{2-6}
&\cite{TreGuiRie:09}&constructive&$K\ge 2$,&$d_k=1$,&$\mnc{N}{\ml}{k}=M_k=K-1,\;\;\forall k$& \\
\cline{2-6}
&\cite{GomCadJaf:08,PetHea:09,PetHea:11,GomCadJaf:11}&iterative& \multicolumn{3}{c}{general configuration}
\\
\hlinew{0.8pt}
\end{tabular}
}
\end{table}

Table~\ref{tab:feasibility} and~\ref{tab:algorithm} outline the contribution of existing works on IA (i.e., with $\tilde{K}=K$, $\mathcal{A}=\mathcal{A}_{\mathrm{all}}$) feasibility analysis and transceiver design.
From these tables, it can be seen that IA feasibility conditions are determined for special configurations, and constructive IA transceiver design algorithms are also only applicable to special cases.
Although existing iterative IA transceiver design algorithms apply to general configurations, they may not converge to a global optimum.
In other words, the outputs of iterative algorithms may not be solutions of the IA problem.
In this paper, we will determine the GIA feasibility conditions and develop algorithms that solve GIA problems for networks with general configuration and alignment sets.

\section{Preliminaries}
\label{sec:background}

In this section, we will outline the mathematical approaches adopted in
the existing theoretical works on IA and illustrate the remaining technical challenges.
Then the notion of algebraic independence will be introduced, which is the most important mathematical concept adopted in this work.

\subsection{Challenge in IA Feasibility Analysis}
\label{sec:pre_f}
\label{sec:previousIA}
\begin{figure}[t] \centering
\includegraphics[scale=0.5]{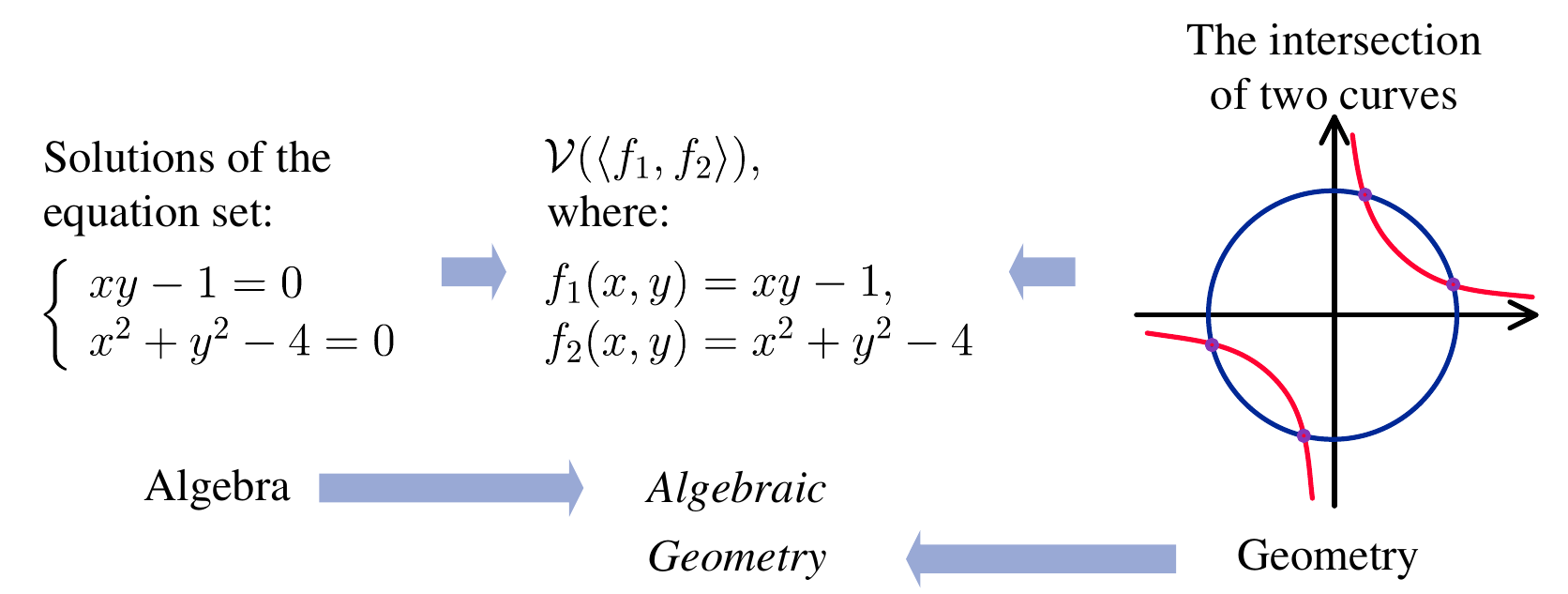}
\caption {An illustration of the correspondence of Algebra, Geometry and \emph{Algebraic geometry}, where $\mathcal{V}(\langle f_1,f_2\rangle)$ denotes the vanishing set of the ideal generated by $\langle f_1,f_2\rangle$ \cite[Def. 1, Section 1.4]{CoxLitShe:06}.}
\label{fig_AG_idea}
\end{figure}
There is an inherent connection between the feasibility of a set of polynomial equations and algebraic geometry \cite{Har:B92}, as illustrated in Fig.~\ref{fig_AG_idea}.
As a result, several prior works on IA feasibility analysis convert the IA problem into a polynomial form and then adopt tools from algebraic geometry. In fact, Problem~\ref{pro:GIA} can be converted to the following polynomial form:\footnote{This statement will be proved formally in Theorem~\ref{thm:F2AI}}

\begin{Prob}[Polynomial Form of GIA Transceiver Design]\label{pro:IA_poly} Design
$\tilde{\mathbf{U}}_{k}\in\mathbb{C}^{(\mnc{N}{\ml}{k}-d_k)\times d_k}$, $\tilde{\mathbf{V}}_{j}\in\mathbb{C}^{(M_j-d_j)\times d_j}$ such that:
\begin{eqnarray}
&&\hspace{-10mm}f_{kjpq}(\{\tilde{\mathbf{U}}_{k},\tilde{\mathbf{V}}_{j}\})\nonumber
\\&\triangleq&\!\! \tilde{\mathbf{u}}^\mathrm{H}_{k}(p)\MNC{H}{\ml}{kj}(d_k\!+\!1:\mnc{N}{\ml}{k},q)
+\!\MNC{H}{\ml}{kj}(p,d_j\!+\!1:M_j)\tilde{\mathbf{v}}_{j}(q)\nonumber
\\ \!\!&&\!\!+\tilde{\mathbf{u}}^\mathrm{H}_{k}(p)
\MNC{H}{\ml}{kj}(d_k\!+\!1:\mnc{N}{\ml}{k}, d_j\!+\!1:M_j)\tilde{\mathbf{v}}_{j}(q)
\label{eqn:czero_poly}
\\\nonumber\!\!&=&\!\!-h_{kj}(p,q),
\end{eqnarray}
where $k,j\in\mathcal{A}$, $p\in\{1,\ldots,d_k\}$, $q\in\{1,\ldots,d_j\}$, $\tilde{\mathbf{u}}_{k}(q)$, and $\tilde{\mathbf{v}}_{j}(q)$ represent the $q$-th column of $\tilde{\mathbf{U}}_{k}$ and $\tilde{\mathbf{V}}_{j}$, respectively. $h_{kj}(p,q)$ is the element in the $p$-th row and $q$-th column of $\MNC{H}{\ml}{kj}$, and $\MNC{H}{\ml}{kj}(p:p',q:q')$ represents the submatrix intersected by $p$ to $p'$-th rows and $q$ to $q'$-th columns of $\MNC{H}{\ml}{kj}$.~\hfill~\IEEEQED\end{Prob}

\setlength{\fboxrule}{0.4pt}
\setlength{\fboxsep}{2mm}
\vspace{3mm}\noindent
\framebox{
\parbox{8.3cm}{
\textbf{Challenge of Nonlinearity}

In the polynomials $f_{kjpq}$ defined above, there are second order terms, i.e., $\tilde{\mathbf{u}}^\mathrm{H}_{k}(p)
\MNC{H}{\ml}{kj}(d_k\!+\!1$ $:\mnc{N}{\ml}{k}, d_j\!+\!1:M_j)\tilde{\mathbf{v}}_{j}(q)$. The presence of these second order terms makes it difficult to analyze the feasible region of Problem~\ref{pro:IA_poly}.
This is because there are very few systematic tools that address the solvability issue of a set of nonlinear polynomial equations.
}}
\vspace{3mm}

\subsection{Challenge in IA Transceiver Design}
\label{sec:pre_a}
Existing IA transceiver design algorithms can be classified into two categories: constructive algorithms and iterative algorithms.
The constructive algorithms design transceivers according to some closed-form functions of the channel states.
However, as illustrated in Table~\ref{tab:algorithm}, these algorithms only apply to limited configurations.

Iterative algorithms are applicable to networks with a general configuration. The most influential iterative algorithm was proposed in \cite{GomCadJaf:08} and \cite{PetHea:09}.\footnote{There
are some differences between the algorithms proposed in \cite{GomCadJaf:08}
and \cite{PetHea:09}. However, the structure and the idea of these two algorithms
are similar.}
This algorithm searches for the IA solution by exploiting the uplink and downlink reciprocity and alternatively updates precoders and decoders in the following problem.

\begin{Prob}[Interference Minimization]\label{pro:IA_opt_s}
\begin{alignat}{3}
&\underset{\mathbf{V}_j,\MNC{U}{\ml}{k}}{\mbox{minimize}} &&\hgap\sum_{k=1}^K\sum_{j=1,\neq
k}^K\frac{P_j}{d_j}\mathrm{trace}\left(\mathbf{V}^\mathrm{H}_j\mathbf{H}^\mathrm{H}_{kj}\MNC{U}{\ml}{k}
(\MNC{U}{\ml}{k})^\mathrm{H}\mathbf{H}_{kj}\mathbf{V}_j\right)\label{eqn:leakage_s1}
\\
&\mbox{subject to} &&\hgap\mathbf{V}^\mathrm{H}_j\mathbf{V}_j=\mathbf{I}, \qquad(\MNC{U}{\ml}{k})^\mathrm{H}\MNC{U}{\ml}{k}=\mathbf{I},
\quad \forall k,j.\label{eqn:unitary_s1}
\end{alignat}\end{Prob}

Although widely adopted in the literature, the alternative minimization algorithm converges to a local optimum.
In other words, it may not be able to cancel all interference even in IA feasible regions.\footnote{That having been said, from the extensive numerical tests in Section~\ref{sec:sim}, we tend to believe that the algorithm proposed in \cite{GomCadJaf:08} converges to a \emph{global} optimum when IA is feasible. However, this conjecture is not proved in the literature.}
The convergence issue is challenging because of the non-convexity challenge elaborated below.

\vspace{3mm}\noindent
\framebox{
\parbox{8.3cm}{
\textbf{Challenge of Non-convexity}
\begin{itemize}
\item[\hspace{-6mm}(1)] The objective function \eqref{eqn:leakage_s1} is not a convex function of the optimization variables ${\mathbf{V}_j,\MNC{U}{\ml}{k}}$;
\item[\hspace{-6mm}(2)] The policy space defined by \eqref{eqn:unitary_s1} is non-convex.
\end{itemize}
}}
\vspace{3mm}

\subsection{Introduction to Algebraic Independence}
\label{sec:algebraic}
To overcome the nonlinearity and non-convexity challenges in the IA problem, a theoretical framework will be developed based on one of the key notions in algebraic geometry, i.e., \emph{algebraic independence}. In this section, the definition of algebraic independence will be introduced and intuitions associated with the notion will be highlighted.

First recall linear independence. Let  $\mathcal{K}$ be
a field, then the standard definition of linear independence is given by:
\begin{Def}[Linear Independence (Form I)]\label{def:LI1} Vectors $\mathbf{a}_l\in\mathcal{K}^S$, $l\in\{1,\ldots,L\}$  are linearly independent iff.
$\sum_{l=1}^L k_l\mathbf{a}_l\neq \mathbf{0}$, $\forall [k_1,\ldots,k_L]\neq \mathbf{0},\in\mathcal{K}^L$.~\hfill~\IEEEQED
\end{Def}

In fact, Definition~\ref{def:LI1} can be transformed to the following equivalent form, which involves linear functions:

\begin{Def}[Linear Independence (From II)]\label{def:LI2} Define \ul{linear functions} $f_l=\sum_{s=1}^{S}$ $a_l(s)x_s$, $l\in\{1,\ldots,L\}$, where $a_l(s)$ is the $s$-th element of $\mathbf{a}_l$. Coefficient vectors $\{\mathbf{a}_l\}$ are linearly independent iff.
$G(f_1,\ldots,f_L)\not\equiv 0$, $\forall$ non-zero \ul{linear function} $G$.~\hfill~\IEEEQED
\end{Def}

With Definition~\ref{def:LI2}, we are ready to introduce algebraic independence. In fact, one just need to replace ``linear function" by ``polynomial" in Definition~\ref{def:LI2} to arrive at the definition for algebraic independence:

\begin{Def}[Algebraic Independence]\label{def:AI} \ul{Polynomials} $f_l\in\mathcal{K}(x_1,\ldots,x_S)$, $l\in\{1,\ldots,L\}$, are algebraically independent
iff. $G(f_1,\ldots,f_L)\not\equiv 0$, $\forall$ non-zero \ul{polynomial} $G\in\mathcal{K}(z_1,\ldots,z_L)$.~\hfill~\IEEEQED
\end{Def}

\begin{Remark}[Linear and Algebraic Independence]
\label{remark:LIAI}The underlined parts in Definition~\ref{def:LI2} and~\ref{def:AI} highlight that algebraic independence is an extension of linear independence.
In the light of this information, it is reasonable to guess the properties of algebraic independence based on those of linear independence.
For instance, if a statement holds conditional on linear independence, it is possible that a similar statement also holds conditional on algebraic independence.
As will be illustrated in Remark~\ref{remark:Intuition}, this intuition does help to construct a unified algebraic framework for both GIA feasibility analysis and algorithm design.
~\hfill~\IEEEQED
\end{Remark}

\section{Feasibility Conditions and Transceiver Design}
\label{sec:result}
In this section, the main theoretical results on the GIA feasibility analysis and transceiver design are proposed and proved.
First, an algebraic framework is established, which shows the (almost sure) equivalence of 1) feasibility of Problem~\ref{pro:GIA}, 2) algebraic independence of $\{f_{kjpq}\}$ defined in \eqref{eqn:czero_poly}, 3) linear independence of the coefficient vectors of the first order terms in $\{f_{kjpq}\}$, and 4) full rankness of the Jacobian matrix of $\{f_{kjpq}\}$.
Based on this framework, a necessary and sufficient feasibility condition of the GIA problem and design algorithms will be given to solve the GIA problem.
\subsection{Mathematical Framework}
\label{subsec:frame}
\begin{figure}[t] \centering
\vspace{-2mm}
\includegraphics[scale=0.5]{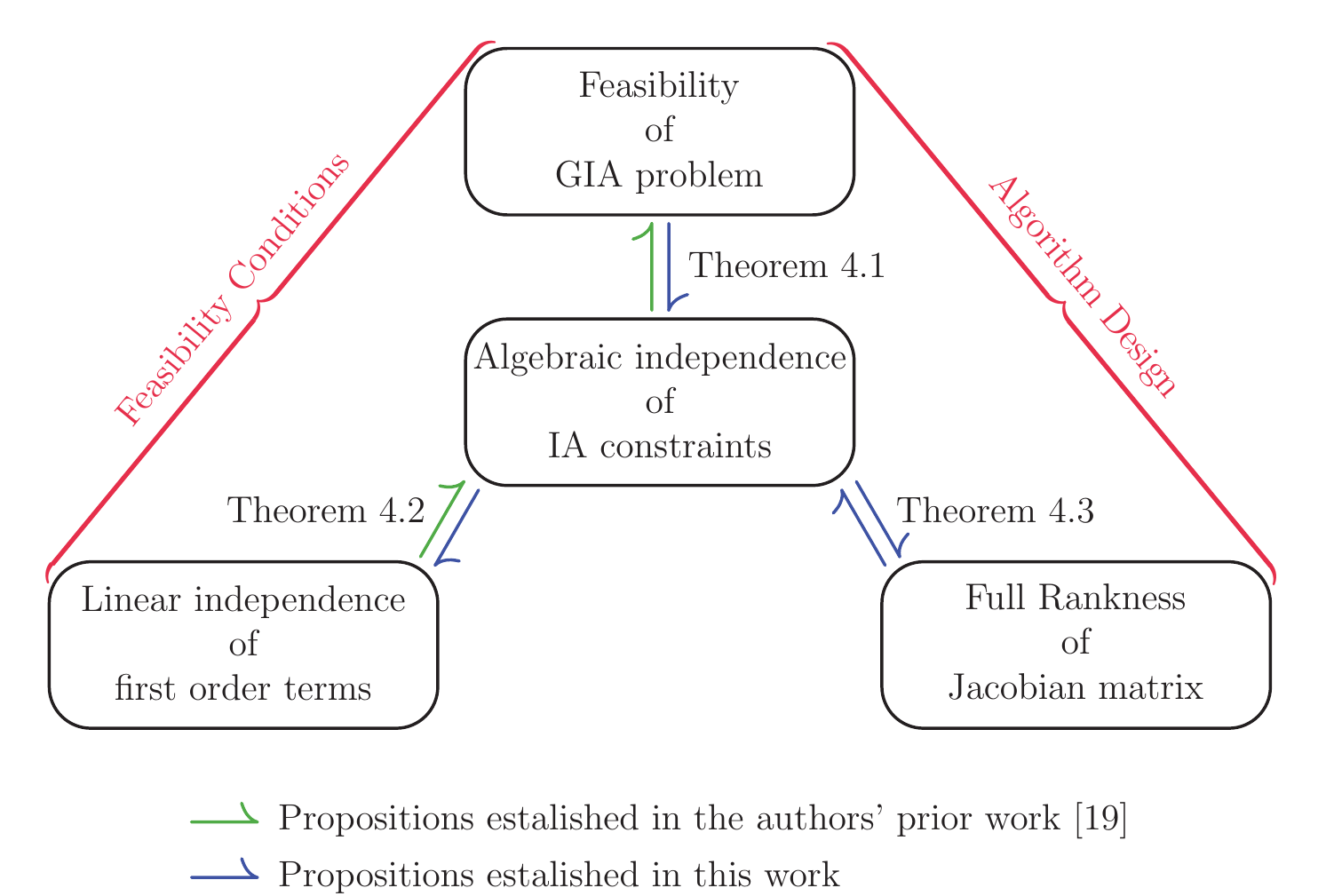}
\caption {Outline of the algebraic framework for the GIA problem.}
\label{fig_Framework}
\end{figure}

We will first define the coefficient matrix of the first order terms of GIA constraints, then list the three theorems that construct the algebraic framework outlined in Fig.~\ref{fig_Framework},
and finally elaborate the intuition of these theorems by showing their counterparts in linear algebra.

\begin{figure*}[t] \centering
\vspace{-2mm}
\includegraphics[scale=0.72]{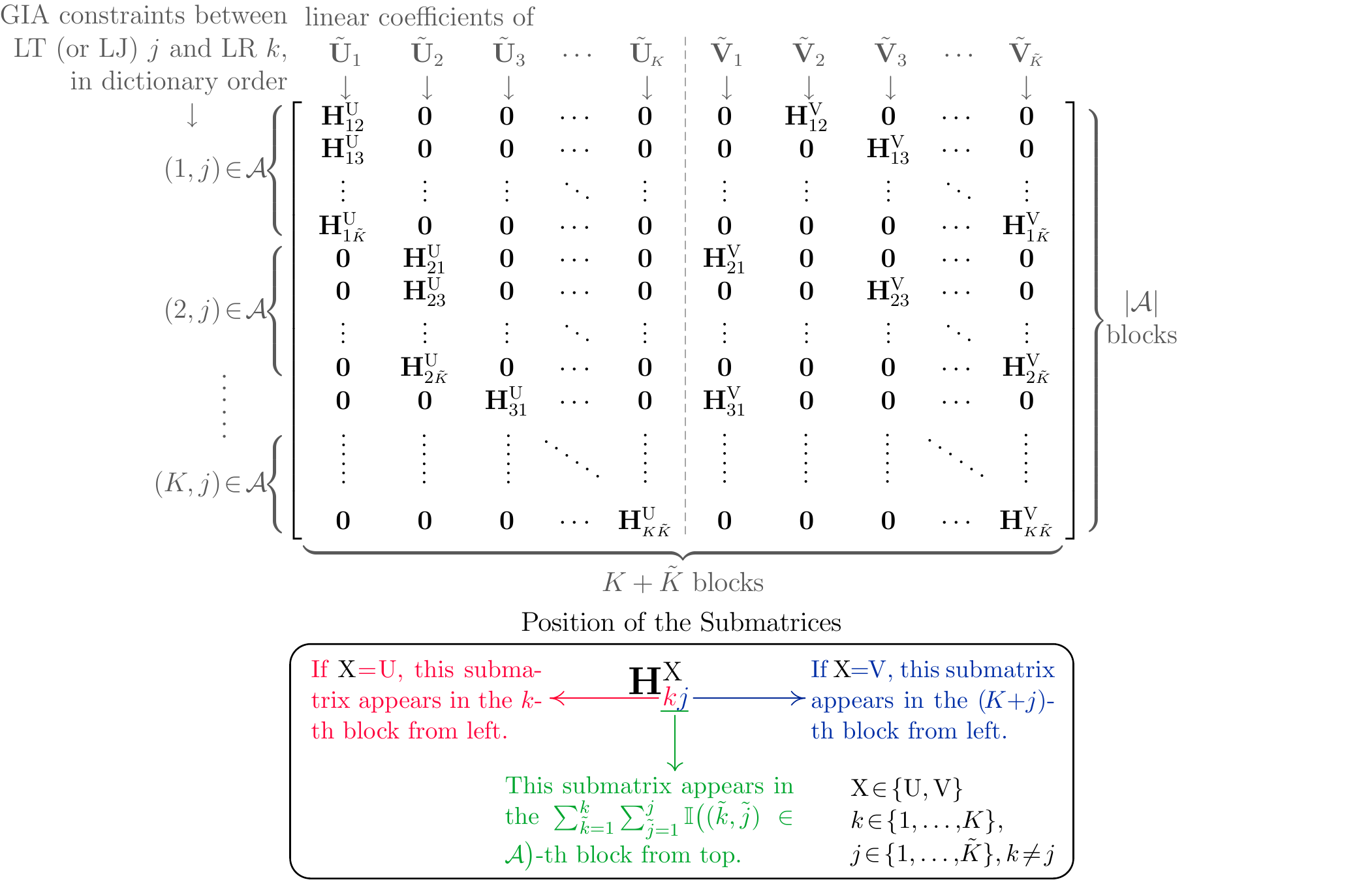}
\caption {The matrix scattered by the coefficient vectors of the linear terms in the polynomial form of the GIA constraints. For clear representation,   $\mathcal{A}$ is set to be equal to $\mathcal{A}_{\mathrm{all}}$ in the figure. When $\mathcal{A}\subset\mathcal{A}_{\mathrm{all}}$, part of the rows will not appear. The zero matrices which appear on the same block row with $\mathbf{H}^{\mathrm{U}}_{kj}$ and $\mathbf{H}^{\mathrm{V}}_{kj}$  have $d_kd_j$ rows. The zero matrices which appear on the same block column with $\mathbf{H}^{\mathrm{U}}_{kj}$ or $\mathbf{H}^{\mathrm{V}}_{kj}$ have $d_k(\mnc{N}{\ml}{k}-d_k)$ and   $d_j(M_j-d_j)$ columns, respectively.}
\label{fig_Hall}
\end{figure*}

Define $\mathbf{H}_{\mathrm{all}}$ as the matrix aggregated by the coefficient vectors of the first order terms in $\{f_{kjpq}\}$.
The structure of $\mathbf{H}_{\mathrm{all}}$ is described in Fig.~\ref{fig_Hall}, where the submatrices $\mathbf{H}^{\mathrm{U}}_{kj}\in\mathbb{C}^{(d_kd_j)\!\times\!(d_k\!(\mnc{N}{\ml}{k}\!-\!d_k))}$ and $\mathbf{H}^{\mathrm{V}}_{kj}\in\mathbb{C}^{(d_kd_j)\!\times\!(d_j\!(M_j\!-\!d_j))}$ are defined by
\begin{eqnarray}
&&\hspace{-11mm}\mathbf{H}^{\mathrm{U}}_{kj}=\nonumber\\
&&\hspace{-8mm}\mathrm{diag}[d_k]\!\!\left(\!\!\begin{array}{*{3}{c@{\,}}c}
h_{kj}(\!d_k\!\!+\!\!1,1),&h_{kj}(\!d_k\!\!+\!\!2,1),
&\!\cdots\!,&h_{kj}(\!\mnc{N}{\ml}{k},1)\\
h_{kj}(\!d_k\!\!+\!\!1,2),&h_{kj}(\!d_k\!\!+\!\!2,2),
&\!\cdots\!,&h_{kj}(\!\mnc{N}{\ml}{k},2)\\
\vdots&\vdots&\ddots&\vdots\\
h_{kj}(\!d_k\!\!+\!\!1, d_j\!),&h_{kj}(\!d_k\!\!+\!\!2, d_j\!),
&\!\cdots\!,&h_{kj}(\!\mnc{N}{\ml}{k}, d_j\!)
\end{array}\!\!\right)\label{eqn:hu}\\
&&\hspace{-11mm}\mathbf{H}^{\mathrm{V}}_{kj}=\nonumber
\\&&\hspace{-8mm}\left[\!\!
\begin{array}{*{4}{c@{\,}}c}
\mathrm{diag}[d_j]\big(&\!\!
h_{kj}(\!1, d_j\!\!+\!\!1\!),&h_{kj}(\!1, d_j\!\!+\!\!2\!),&\!\cdots\!,&h_{kj}(\!1,\!M_j\!)\big)\\
\mathrm{diag}[d_j]\big(&\!\!
h_{kj}(\!2, d_j\!\!+\!\!1\!),&h_{kj}(\!2, d_j\!\!+\!\!2\!),&\!\cdots\!,&h_{kj}(\!2,\!M_j\!)\big)\\
\multicolumn{5}{c}{\cdots\cdots}\\
\mathrm{diag}[d_j]\big(&\!\!
h_{kj}(\!d_k, d_j\!\!+\!\!1\!),&h_{kj}(\!d_k, d_j\!\!+\!\!2\!),&\!\cdots\!,&h_{kj}(\!d_k,\!M_j\!)\big)
\end{array}
\!\!\!\right]\label{eqn:hv}
\end{eqnarray}
where $h_{kj}(p,q)$ denotes the element in the $p$-th row and $q$-th column of $\MNC{H}{\ml}{kj}$, $k\neq j, k\in\{1,\ldots,K\}, j\in\{1,\ldots,\tilde{K}\}$. Note that the coefficient vectors of the first order terms in $\{f_{kjpq}\}$ are linearly independent iff. $\mathbf{H}_{\mathrm{all}}$ is full row-rank.

The following three theorems construct the algebraic framework for GIA feasibility analysis and algorithm design.

\begin{Thm}[Equivalence of Feasibility and Algebraic Independence]\label{thm:F2AI} Under a network configuration $\chi$, Problem~\ref{pro:GIA} has solutions almost surely\footnote{In this paper, ``almost surely" means ``with probability 1."} iff. the polynomials $\{f_{kjpq}\}$ defined in \eqref{eqn:czero_poly} are algebraically independent. The solution of Problem~\ref{pro:GIA} can be obtained by first solving Problem~\ref{pro:IA_poly} and then constructing transceivers $\{\MNC{U}{\ml}{k},\mathbf{V}_j\}$ via \eqref{eqn:tildeuv}:
\begin{eqnarray}\MNC{U}{\ml}{k}=\left[\!\!\!\begin{array}{c}\mathbf{I}_{d_k\times d_k}\\ \tilde{\mathbf{U}}_{k}\end{array}\!\!\!\right],\; \mathbf{V}_j=\left[\!\!\!\begin{array}{c}\mathbf{I}_{d_j\times d_j}\\ \tilde{\mathbf{V}}_{j}\end{array}\!\!\!\right].
\label{eqn:tildeuv}
\end{eqnarray}
\end{Thm}

\begin{IEEEproof} Please refer to Appendix~\ref{pf_thm:F2AI} for the proof.
\end{IEEEproof}

\begin{Thm}[Equivalence of Algebraic Independence and Linear Independence]\label{thm:AI2LI} Under a network configuration $\chi$, matrix $\mathbf{H}_{\mathrm{all}}$ (defined in Fig.~\ref{fig_Hall}) is either full row-rank almost surely or always row-rank deficient. In the first case, the polynomials $\{f_{kjpq}\}$ defined in \eqref{eqn:czero_poly} are almost surely algebraically independent. Otherwise,  $\{f_{kjpq}\}$ are algebraically dependent.
\end{Thm}

\begin{IEEEproof} Please refer to Appendix~\ref{pf_thm:AI2LI} for the proof.
\end{IEEEproof}

\begin{Thm}[Equivalence of Algebraic Independence and Nonsingularity of Jacobian Matrix]\label{thm:AI2J} The polynomials $\{f_{kjpq}\}$ defined in \eqref{eqn:czero_poly} are algebraically independent iff. the Jacobian matrix $\mathbf{J}_{\mathbf{x}}(\{f_{kjpq}\})$ is full row-rank on a dense and open subset of $\mathbb{C}^V$, where $V=\sum_{k=1}^K d_k(\mnc{N}{\ml}{k}-d_k)+\sum_{j=1}^{\tilde{K}} d_j(M_j-d_j)$.
\end{Thm}

\begin{IEEEproof} Please refer to Appendix~\ref{pf_thm:AI2J} for the proof.
\end{IEEEproof}

\begin{Remark}[Intuition from Linear Independence]
\label{remark:Intuition}
To interpret the algebraic framework outlined in Fig.~\ref{fig_Framework}, consider a set of linear functions:
\begin{eqnarray}
f_l(x_1,\ldots,x_S)=\sum_{s=1}^{S}a_l(s)x_s=\mathbf{a}_l\mathbf{x}, \qquad l\in\{1,\ldots,L\},
\end{eqnarray}
where coefficient vector $\mathbf{a}_l=[a_l(1),\ldots,a_l(S)]\in\mathbb{C}^S$ and variable vector $\mathbf{x}=[x_1,\ldots,x_S]^{\mathrm T}\in\mathbb{C}^S$. Define $\mathbf{A}=\begin{bmatrix}\mathbf{a}_1\vspace{-1.5mm}\\\vdots\vspace{-1.5mm}\\ \mathbf{a}_L\end{bmatrix}$. From linear algebra, the following proposition holds:\vspace{1mm}
\begin{Prop}[Equivalence of Linear Independence and Feasibility]\label{prop:LI2F} Consider a
 vector $\mathbf{b}=[b_1,\ldots,b_L]^{\mathrm T}$ whose elements are independent random variables drawn from continuous distribution.
Then linear equation set $f_l=b_l$, $l\in\{1,\ldots,L\}$, i.e., $\mathbf{A}\mathbf{x}=\mathbf{b}$ has solutions iff. vectors $\mathbf{a}_1,\ldots,\mathbf{a}_L$ are linearly independent.
\end{Prop}

Furthermore, for any vector $\mathbf{x}\in\mathbb{C}^S$, the Jacobian matrix is
\begin{eqnarray}
\mathbf{J}_\mathbf{x}(f_1,\ldots,f_L)
=\begin{bmatrix}
\frac{\partial f_1}{\partial x_1} &\cdots&\frac{\partial f_1}{\partial x_S}\\
\vdots&\ddots&\vdots\\
\frac{\partial f_L}{\partial x_1} &\cdots&\frac{\partial f_L}{\partial x_S}\\
\end{bmatrix}
= \mathbf{A}.
\end{eqnarray}
Hence, the following proposition is also true:
\begin{Prop}[Equivalence of Linear Independence and Nonsingularity of Jacobian Matrix]
\label{prop:LI2J}
The coefficient vectors of $f_1,\ldots,f_L$ are linearly independent iff. the Jacobian matrix $\mathbf{J}_\mathbf{x}(f_1,\ldots,f_L)$ is full row-rank for any $\mathbf{x}\in\mathbb{C}^S$.
\end{Prop}

By comparing Proposition~\ref{prop:LI2F} and~\ref{prop:LI2J} with Theorem~\ref{thm:F2AI} and~\ref{thm:AI2J}, it can be seen that linear independence and algebraic independence play a similar role in these statements.
This fact fits the insight illustrated in Remark~\ref{remark:LIAI}.
Actually, in the authors' previous work \cite[Lem. 3.1]{RuaLauWin:J13}, it was shown that if the coefficient vectors of the first order terms of a set of polynomials are  {\em linearly} independent, then these polynomials are {\em algebraically} independent.
The inverse proposition of \cite[Lem. 3.1]{RuaLauWin:J13} is not true for general polynomials.
Yet, in this paper, by exploiting the special structure of the polynomials defined in \eqref{eqn:czero_poly}, the inverse proposition for GIA problems has been proved and hence Theorem~\ref{thm:AI2LI} is obtained.~\hfill~\IEEEQED
\end{Remark}
\subsection{Feasibility Conditions}
\vspace{3mm}\noindent
\framebox{
\parbox{8.3cm}{
\textbf{Solution to the Challenge of Nonlinearity}

Based on the algebraic framework established in Section~\ref{subsec:frame}, we have the following theorem which determines the feasibility condition of GIA.

\begin{Thm}[Necessary and Sufficient Feasibility Condition]\label{thm:feasible}
Problem~\ref{pro:GIA} has solutions almost surely iff. matrix $\mathbf{H}_{\mathrm{all}}$ in Fig.~\ref{fig_Hall} is full row-rank.
\end{Thm}

\begin{IEEEproof} This theorem is an immediate consequence of Theorem~\ref{thm:F2AI} and~\ref{thm:AI2LI}.
\end{IEEEproof}
}}
\vspace{3mm}

With Theorem~\ref{thm:feasible}, there are three propositions illustrating the general trends on GIA feasibility.

\begin{Cor}[Configuration and Alignment Set Dominate GIA Feasibility]\label{cor:confdomin}
Under given network configuration $\chi$ and alignment set $\mathcal{A}$, Problem~\ref{pro:GIA} is either always infeasible or feasible almost surely.
\end{Cor}
\begin{IEEEproof} This corollary is an immediate consequence of Theorem~\ref{thm:feasible} and Lemma~\ref{lem:rank}.
\end{IEEEproof}

\begin{Cor}[Scalability of GIA Feasibility]\label{cor:scaleable} Under given alignment set $\mathcal{A}$, scaling the legitimate network configuration does not affect the GIA feasibility state, i.e., networks with configuration $\chi=\{(cM_1,\ldots,cM_{\tilde{K}}),(c\mnc{N}{\ml}{1},\ldots,
c\mnc{N}{\ml}{K}),$ $(cd_1,\ldots,cd_{\tilde{K}})\}$, $\forall c\in\mathbb{N}$ are either all GIA feasible or all GIA infeasible.
\end{Cor}
\begin{IEEEproof} The proof is similar to that of \cite[Cor. 3.2]{RuaLauWin:J13}. The details are omitted to avoid redundancy.
\end{IEEEproof}

\begin{Remark}[Contributions of Corollary~\ref{cor:confdomin},~\ref{cor:scaleable}]\label{rem:test}
Theorem~\ref{thm:feasible} gives a complete characterization of the feasibility condition of GIA problems.
However, the feasibility condition in Theorem~\ref{thm:feasible} is complicated as it relates to network configuration $\chi$, alignment set $\mathcal{A}$, as well as the instantaneous channel state $\{\MNC{H}{\ml}{kj}\}$.
Corollary~\ref{cor:confdomin} simplifies this condition by showing that with probability 1, the feasible state is determined by configuration $\chi$ and alignment set $\mathcal{A}$.
Corollary~\ref{cor:scaleable} further simplifies this condition by showing that networks with configurations different by a factor share the same feasible state.

One application of the  propositions is an efficient method to check GIA feasibility. To determine if a set of networks with configuration  $\chi=\{(cM_1,\ldots,cM_{\tilde{K}}),(c\mnc{N}{\ml}{1},\ldots,
c\mnc{N}{\ml}{K}),$ $(cd_1,\ldots,cd_{\tilde{K}})\}$, $\forall c\in\mathbb{N}$ is GIA feasible or not:  set $c=1$, randomly generate one channel state, and check if $\mathbf{H}_{\mathrm{all}}$ is full row-rank or not.
~\hfill~\IEEEQED
\end{Remark}

\begin{Cor}[Necessary GIA Feasibility Condition]\label{cor:proper} A  network with configuration $\chi$ and alignment set $\mathcal{A}$ is GIA feasible only if
\begin{eqnarray}
&&\hspace{-8mm}\sum_{j:(k,j)\in\mathcal{A}_{\mathrm{sub}}}\hspace{-3mm}d_j(M_j-d_j)
+\sum_{k:(k,j)\in\mathcal{A}_{\mathrm{sub}}}\hspace{-3mm}d_k(\mnc{N}{\ml}{k}-d_k)\ge\nonumber\\ &&\hspace{-7mm}\sum_{(k,j)\in\mathcal{A}_{\mathrm{sub}}}\hspace{-2mm}d_kd_j,\qquad\forall\mathcal{A}_{\mathrm{sub}}\subseteq \mathcal{A}.\label{eqn:proper0}
\end{eqnarray}
\end{Cor}

\begin{IEEEproof} Denote $\mathbf{H}_{\mathrm{sub}}$ as the submatrix of $\mathbf{H}_{\mathrm{all}}$ that corresponds to $\mathcal{A}_{\mathrm{sub}}$. $\mathbf{H}_{\mathrm{sub}}$ has $\sum_{(k,j)\in\mathcal{A}_{\mathrm{sub}}}\hspace{-1mm}d_kd_j$ rows and $\sum_{j:(k,j)\in\mathcal{A}_{\mathrm{sub}}}\hspace{-1mm}d_j(M_j-d_j)
+\sum_{k:(k,j)\in\mathcal{A}_{\mathrm{sub}}}\hspace{-1mm}d_k(\mnc{N}{\ml}{k}-d_k)$ non-zero columns. Hence, when \eqref{eqn:proper0} does not hold for a certain $\mathcal{A}_{\mathrm{sub}}$, the corresponding $\mathbf{H}_{\mathrm{sub}}$ is row-rank deficient and so is $\mathbf{H}_{\mathrm{all}}$. From Theorem~\ref{thm:feasible}, the network is infeasible. This completes the proof.
\end{IEEEproof}

\begin{Remark}[Properness and Feasibility] In the pioneering work on IA feasibility analysis \cite{YetGouJaf:10}, the authors conjecture that a MIMO interference network is IA feasible only if the network is proper; i.e., the number of variables in transceiver design is no more than the number of IA constraints.
This conjecture was later confirmed by \cite{BreCarTse:11} and \cite{RazGenLuo:12}.
Corollary~\ref{cor:proper} shows that properness is still a necessary feasibility condition for GIA problems.
~\hfill~\IEEEQED
\end{Remark}

In the following,  two corollaries are given which reveal simple insights into how legitimate network configuration $\chi$ and alignment set $\mathcal{A}$ determine the GIA feasibility.

\begin{Cor}[Symmetric Configuration]\label{cor:sym} When 1) network configuration $\chi$ is symmetric, i.e., $d_k=d$, $M_k=M$, and $\mnc{N}{\ml}{k}=N$, $\forall k\in\{1,\ldots,K\}$, with $\min\{M,N\}\ge 2d$; 2) alignment set between the LRs and LTs is $L$\emph{-regular}, i.e., $\sum_{j=1}^K\mathbb{I}\{(k,j)\in\mathcal{A}\} = \sum_{k=1}^K\mathbb{I}((k,j)\in\mathcal{A})=L$, $\forall k,j\in\{1,\ldots,K\}$;  and 3) each LJ chooses the proper number of LRs to coordinate with, i.e., $\sum_{k=1}^K\mathbb{I}((k,j)\in\mathcal{A})\le\left\lfloor\frac{M_j-d_j}{d}\right\rfloor$, $\forall j\in\{K+1,\ldots,\tilde{K}\}$, Problem~\ref{pro:GIA} has solutions almost surely iff. inequality \eqref{eqn:f_sym} is true, where
\begin{eqnarray}M+N-(L+2)d\ge0. \label{eqn:f_sym}
\end{eqnarray}
\end{Cor}

\begin{IEEEproof} Please refer to Appendix~\ref{pf_cor:sym} for the proof.
\end{IEEEproof}

\begin{Cor}[``Divisible" Configuration]\label{cor:div}When the network configuration $\chi$ satisfies 1) $d_k=d$, $\forall k\in\{1,\ldots,\tilde{K}\}$ and 2) $d|\mnc{N}{\ml}{k}$, $\forall k\in\{1,\ldots,K\}$ or $d|M_k$, $\forall k\in\{1,\ldots,\tilde{K}\}$, Problem~\ref{pro:GIA} has solutions almost surely iff. inequality \eqref{eqn:f3_divisor} is satisfied, where\vspace{-5pt}
\begin{eqnarray}
\sum_{j:(k,j)\in\mathcal{A}_{\mathrm{sub}}}\hspace{-2mm}(M_j-d)
+\sum_{k:(k,j)\in\mathcal{A}_{\mathrm{sub}}}\hspace{-2mm}(\mnc{N}{\ml}{k}-d)\ge d|\mathcal{A}_{\mathrm{sub}}|,\label{eqn:f3_divisor}
\end{eqnarray}
$\qquad\forall\mathcal{A}_{\mathrm{sub}}\subseteq
\mathcal{A}$.
\end{Cor}

\begin{IEEEproof} Please refer to Appendix~\ref{pf_cor:div} for the proof.
\end{IEEEproof}

\begin{Remark}[Backward Compatibility to Existing Works]
If one specify the GIA problem to the classical IA problem, i.e., sets $\tilde{K}=K$ and $\mathcal{A}=\mathcal{A}_{\mathrm{all}}$, then Corollary~\ref{cor:sym} and~\ref{cor:div} are reduced to \cite[Cor. 3.3]{RuaLauWin:J13} and \cite[Cor. 3.4]{RuaLauWin:J13}, respectively.
Further noting that \cite[Cor. 3.3]{RuaLauWin:J13} and \cite[Cor. 3.4]{RuaLauWin:J13} extend the feasibility conditions proved in \cite{BreCarTse:11} and \cite{RazGenLuo:12}, respectively, Corollary~\ref{cor:sym} and~\ref{cor:div} are consistent with prior theoretical results on IA feasibility and extend these results to the GIA case.
~\hfill~\IEEEQED
\end{Remark}

\subsection{GIA Transceiver Design}
\begin{figure}[t] \centering
\includegraphics[scale=0.48]{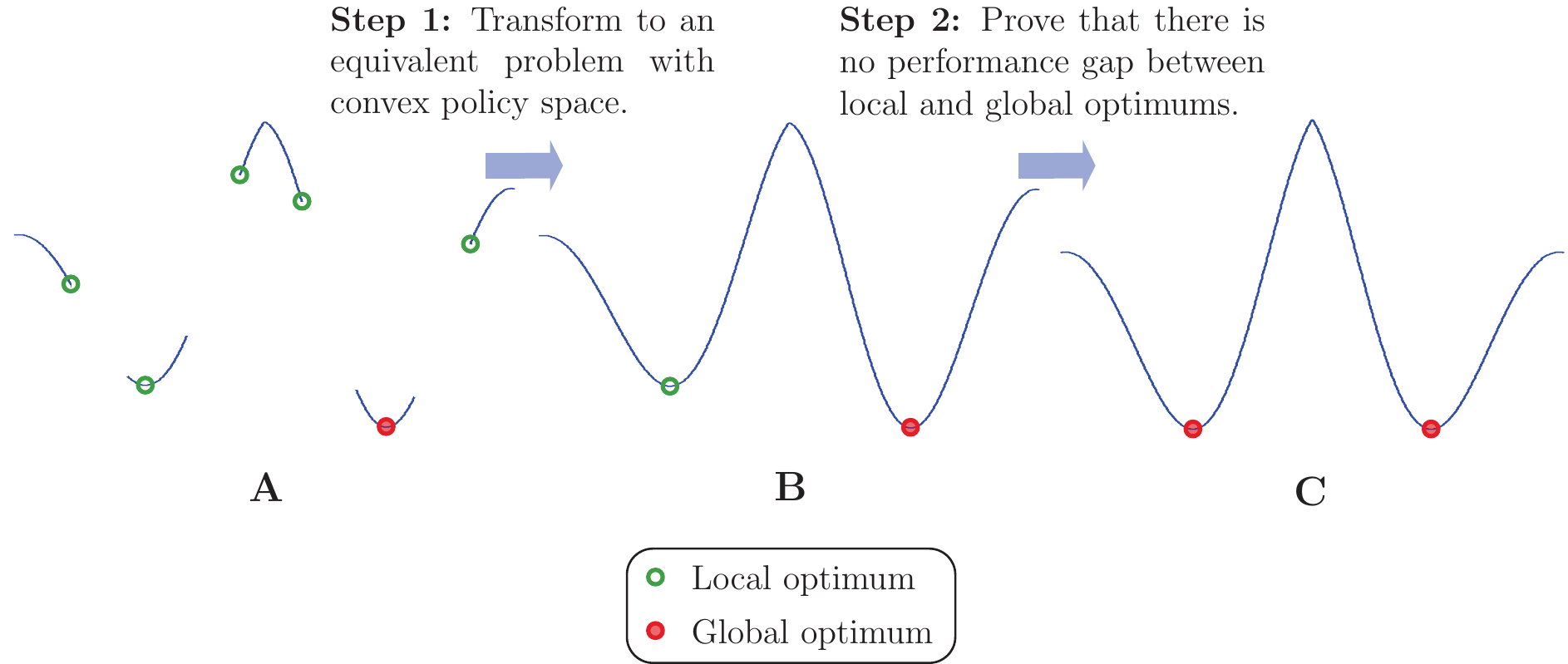}
\caption {An intuitive illustration of how the algebraic framework established in this work enables us to find the global optimum of the interference minimization problem.} \label{fig_outline}
\end{figure}

As illustrated in Section~\ref{sec:pre_a}, IA transceiver design is challenging because
neither the policy space nor the objective function of the interference minimization problem is convex.
Fig.~\ref{fig_outline} gives an intuitive illustration of how this challenge will be overcome.
In the first step, transform the problem to an equivalent one with convex policy space.
In the second step, prove that there is no performance gap between the local and global optimums.
Hence, despite the fact that the objective function is non-convex, the problem can be solved by various local search algorithms.

\vspace{3mm}\noindent
\framebox{
\parbox{8.3cm}{
\textbf{Solution to the Challenge of Non-convexity (Step 1)}

In Problem~\ref{pro:IA_poly}, the policy space is given by $\prod_{k=1}^{K}\mathbb{C}^{(\mnc{N}{\ml}{k}-d_k)\times d_k}\cdot\prod_{j=1}^{\tilde{K}}\mathbb{C}^{(M_j-d_j)\times d_j}$, which is a convex set. Hence, the first step is achieved by Theorem~\ref{thm:F2AI}.
}}
\vspace{3mm}

Then, transform Problem~\ref{pro:IA_poly} to the following optimization problem (Problem~\ref{pro:IA_polyopt}).
Note that Problem~\ref{pro:IA_poly} is solved iff. there exists a solution in Problem~\ref{pro:IA_polyopt} that satisfies $F(\{g_{kjpq}(\tilde{\mathbf{U}}^*_k,\tilde{\mathbf{V}}^*_j)\})$ $=0$.

\begin{Prob}[Optimization Form of GIA Problem]\label{pro:IA_polyopt}
\begin{eqnarray}
\underset{\tilde{\mathbf{U}}_{k}\in\mathbb{C}^{(\mnc{N}{\ml}{k}-d_k)\times d_k}\atop \tilde{\mathbf{V}}_{j}\in\mathbb{C}^{(M_j-d_j)\times d_j}}{\mbox{minimize}}F(\{g_{kjpq}(\tilde{\mathbf{U}}_{k},\tilde{\mathbf{V}}_{j})\}),\label{eqn:czero_opt}
\end{eqnarray}
where $g_{kjpq}=f_{kjpq}+h_{kj}(p,q)$, $f_{kjpq}$ is defined in \eqref{eqn:czero_poly}, $(k,j)\in\mathcal{A}$, $p\in\{1,\ldots,d_k\}$, $q\in\{1,\ldots,d_j\}$, and $F$ is a nonnegative, convex and continuously differentiable function. $F(\{g_{kjpq}(\tilde{\mathbf{U}}_{k},\tilde{\mathbf{V}}_{j})\})=0$ iff. $g_{kjpq}=0$, $\forall k,j,p,q$.
~\hfill~\IEEEQED\end{Prob}

\vspace{3mm}\noindent
\framebox{
\parbox{8.3cm}{
\textbf{Solution to the Challenge of Non-convexity (Step 2)}

The following theorem achieves the second step in Fig.~\ref{fig_outline} by exploiting Theorem~\ref{thm:AI2J}.

\begin{Thm}[No Gap between Local and Global Optimums] When the polynomial form of the GIA problem, i.e., Problem~\ref{pro:IA_poly} is feasible, in Problem~\ref{pro:IA_polyopt}, every local optimum is globally optimal.\label{thm:nogap}
\end{Thm}
\begin{IEEEproof} Please refer to Appendix~\ref{pf_thm:nogap} for the proof.
\end{IEEEproof}
}}
\vspace{3mm}
\begin{Remark}[The Role of Nonsingular Jacobian Matrix]
The full row-rankness of the Jacobian matrix
$\mathbf{J}_{\tilde{\mathbf{U}}_{k},\tilde{\mathbf{V}}_{j}}(\{g_{kjpq}\})$
plays a key role in the proof of Theorem~\ref{thm:nogap}.
To see how it works, consider a polynomial map $G: \mathbb{C}^N\rightarrow \mathbb{C}^M$. At point $\mathbf{x}_0\in\mathbb{C}^N$,\begin{eqnarray}
G(\mathbf{x}_0+\Delta\mathbf{x})=G(\mathbf{x}_0)+\mathbf{J}_\mathbf{x_{0}}(G)
\Delta\mathbf{x}+\mathcal{O}\big(||\Delta\mathbf{x}||^2\big)
\label{eqn:Jac}
.\end{eqnarray}
Consider a neighborhood of $\mathbf{x}_0$ with
$||\Delta\mathbf{x}||\ll 1$ and suppose the Jacobian matrix $\mathbf{J}_{\mathbf{x}_0}(G)$
is full row rank. In this case, the third term on the right-hand-side of \eqref{eqn:Jac} can be ignored compared to the second term, and $\Delta G = G(\mathbf{x}_0+\Delta\mathbf{x})-G(\mathbf{x}_0)=\mathbf{J}_\mathbf{x_{0}}(G)
\Delta\mathbf{x}$ can be {\em any vector} in the neighborhood of $\mathbf{0}$.

Cascade $G$ with a convex function $F: \mathbb{C}^M\rightarrow \mathbb{R}$, and suppose $\mathbf{x}_0$ is a local optimum of $F(G(\mathbf{x}))$. Then from the definition of local optimum and the property of $\Delta G$  just obtained,  $F(G(\mathbf{x}_0)+\Delta G)\ge F(G(\mathbf{x}_0))$ for any vector $\Delta G$ in the neighbourhood of $\mathbf{0}$. This implies $\mathbf{y}_0=G(\mathbf{x}_0)$ is a local optimum of $F$. Since $F$ is convex, $F(\mathbf{y}_0)$ must also be a global optimum of $F$ and therefore   $\mathbf{x}_0$ is a global optimum of $F(G(\mathbf{x}))$.

For Theorem~\ref{thm:nogap}, there is a weaker condition on the nonsingularity of the Jacobian matrix, i.e., full row-rank on a dense open subset. Yet, by imposing a stronger condition on the form of $F$, i.e., being continuously differentiable, the proof can be completed.
~\hfill~\IEEEQED
\end{Remark}

\begin{figure}[t] \centering
\includegraphics[scale=0.55]{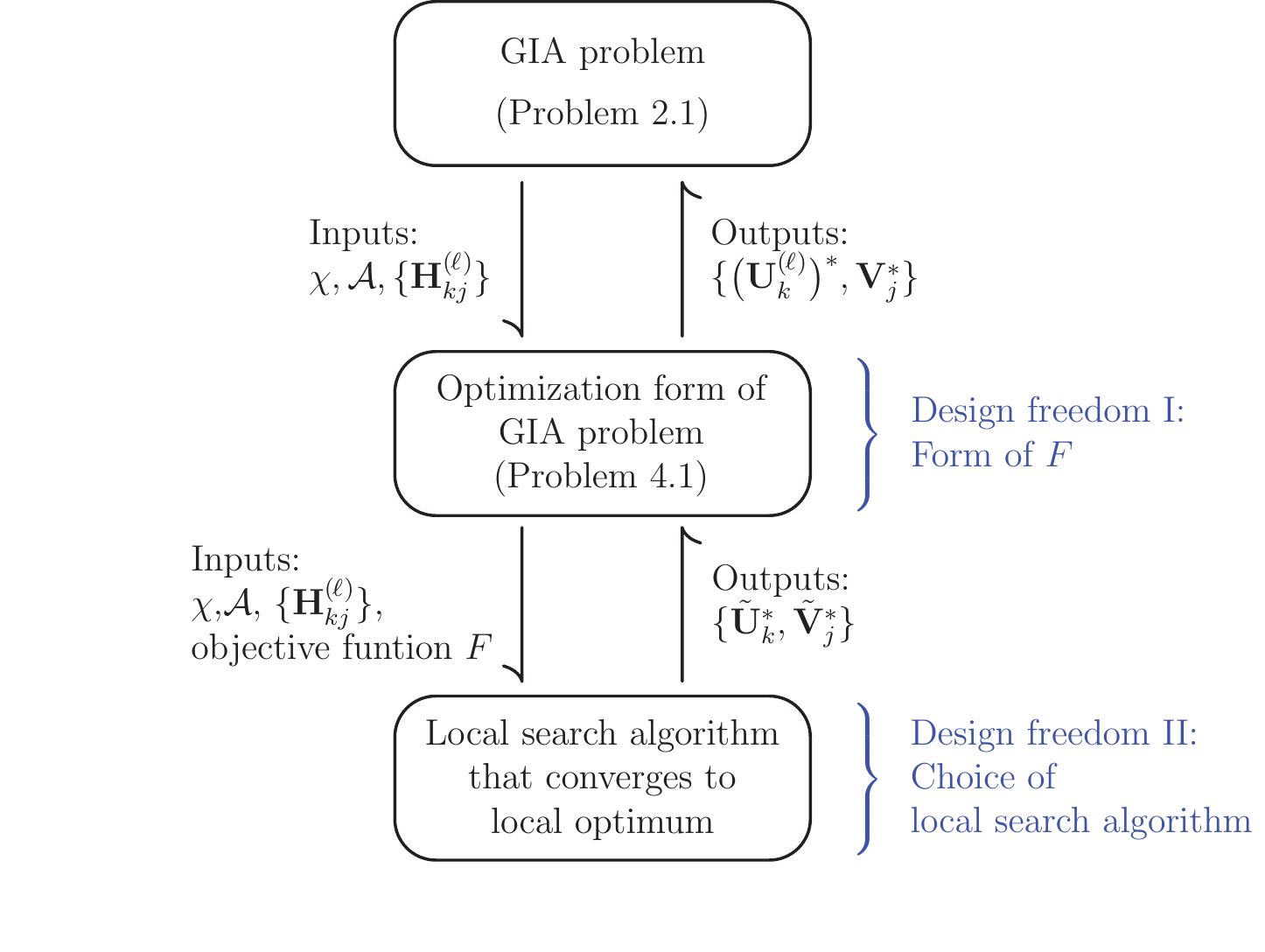}
\caption {Outline of GIA transceiver design algorithms based on Theorem~\ref{thm:nogap}.} \label{fig_algorithm}
\end{figure}

\begin{Remark}[Theoretical Basis for GIA Transceiver Design] As illustrated in Fig.~\ref{fig_algorithm}, based on Theorem~\ref{thm:nogap}, one can generate a set of algorithms that solve the GIA transceiver design problem.
Moreover, the freedom in designing the specific form of $F$ and choosing local search algorithms can be exploited to improve algorithm performances, such as message overhead, convergence speed and throughput.
Hence, Theorem~\ref{thm:nogap} sets up a theoretical basis to design and improve GIA transceiver design algorithms.
~\hfill~\IEEEQED\end{Remark}

\begin{Remark}[Consistency with Existing Theoretical Result]
As illustrated in \cite{RazGenLuo:12a}, IA transceiver design is highly challenging
because ``it is impossible to propose an algorithm that converges to an aligned solution in
polynomial time for each system configuration and for {\em any} set
of channel matrices."
On the other hand, the authors of \cite{RazGenLuo:12a} also predicted that ``there might still exist a polynomial time algorithm that can solve the problem ... with high probability (e.g., for almost all channel coefficients)."
Noting that the polynomial form of the GIA transceiver design problem, i.e., Problem~\ref{pro:IA_poly} is equivalent to the original GIA transceiver design problem, i.e., Problem~\ref{pro:GIA} for almost all channel coefficients, the algorithms outlined in Fig.~\ref{fig_algorithm} solve the original GIA transceiver design problem almost surely.
In this sense, this result confirms the prediction made in \cite{RazGenLuo:12a}.
~\hfill~\IEEEQED
\end{Remark}

As an illustration, one specific algorithm will be presented to achieve GIA. Let $F(\{x_i\})
=\sum_{i}x_ix_i^\mathrm{H}$; then Problem~\ref{pro:IA_polyopt} can be rewritten as the follows:

\begin{Prob}[Reformed Interference Minimization]\label{pro:IA_opt2}
\begin{eqnarray}
\hspace{-3mm}&\underset{\tilde{\mathbf{V}}_j,\tilde{\mathbf{U}}_k}{\mbox{minimize}}&
\sum_{k=1}^K\sum_{j:(k,j)\in\mathcal{A}\ }||\mathbf{U}^\mathrm{H}_k\mathbf{H}_{kj}\mathbf{V}_j||^2_{\mathrm F}\label{eqn:leakage_2}
\\\nonumber\hspace{-3mm}&\mbox{subject to}& \mbox{Eq. } \eqref{eqn:tildeuv}
\end{eqnarray}
~\hfill~\IEEEQED\end{Prob}

The following algorithm solves Problem~\ref{pro:IA_opt2}:
\begin{Alg}[GIA Transceiver Design]\label{alg:IA}
\begin{itemize}
\item{\bf Step 1 Initialization :} Randomly generate
$\tilde{\mathbf{V}}_{j}$, $j\in\{1,\ldots,\tilde{K}\}$.
\item{\bf Step 2 Minimize interference leakage at the receiver
side:} At LR $k$, update $\tilde{\mathbf{U}}_{k}$:
\begin{eqnarray}
\tilde{\mathbf{U}}_{k}=-\left(\mathbf{B}_{k}\mathbf{A}_{k}^\sharp\right)^\mathrm{H}\label{eqn:alg_U}
,\end{eqnarray}
where

$\mathbf{X}_{k}=[\mathbf{X}_{kj_1},\ldots,
\mathbf{X}_{kj_T}]$, $\{j_1,\ldots,j_T\}=\{j:(k,j)\in\mathcal{A}\}$, $\mathbf{X}\in\{\mathbf{A},\mathbf{B}\}$,

$\mathbf{A}_{kj}=\mathbf{H}_{kj}(d_k\!+\!1:\mnc{N}{\ml}{k},1:d_j)+
\mathbf{H}_{kj}(d_k\!+\!1:\mnc{N}{\ml}{k},d_j\!+\!1:M_j)\tilde{\mathbf{V}}_{j}$, and

$\mathbf{B}_{kj}=\mathbf{H}_{kj}(1:d_k,1:d_j)+
\mathbf{H}_{kj}(1:d_k,d_j\!+\!1:M_j)\tilde{\mathbf{V}}_{j}$.

\item{\bf Step 3 Minimize interference leakage at the transmitter side:}
At LT (LJ) $j$, update $\tilde{\mathbf{V}}_{j}$:
\begin{eqnarray}
\tilde{\mathbf{V}}_{j}=-\mathbf{C}_{j}^\sharp\mathbf{D}_{j}\label{eqn:alg_V}
,\end{eqnarray}
where

$\mathbf{X}_{\bar{j}}=\begin{bmatrix}\mathbf{X}_{k_1j}\vspace{-1mm}\\\vdots\vspace{-1mm}\\
\mathbf{X}_{k_Rj}\end{bmatrix}$, $\{k_1,\ldots,k_R\}=\{k:(k,j)\in\mathcal{A}\}$, $\mathbf{X}\in\{\mathbf{C},\mathbf{D}\}$,

$\mathbf{C}_{kj}=\mathbf{H}_{kj}(1:d_k,d_j\!+\!1:M_j)+
\tilde{\mathbf{U}}^\mathrm{H}_{k}\mathbf{H}_{kj}(d_k\!+\!1:\mnc{N}{\ml}{k},d_j\!+\!1:M_j)$, and

$\mathbf{D}_{kj}=\mathbf{H}_{kj}(1:d_k,1:d_j)+
\tilde{\mathbf{U}}^\mathrm{H}_{k}\mathbf{H}_{kj}(d_k\!+\!1:\mnc{N}{\ml}{k},1:d_j)$.

\item Repeat Step 2 and 3 until $\tilde{\mathbf{V}}_{j}$ and $\tilde{\mathbf{U}}_{k}$
converge. Substitute in \eqref{eqn:tildeuv} and obtain
$\{\mathbf{V}^{*}_{j},{\MNC{U}{\ml}{k}}^*\}$.~\hfill~\IEEEQED
\end{itemize}
\end{Alg}

\begin{Cor}[Convergence of Algorithm~\ref{alg:IA}]\label{thm:opt} Algorithm~\ref{alg:IA} always converges. Moreover, when IA is feasible, the output of Algorithm~\ref{alg:IA}, i.e., $\{\mathbf{V}^{*}_{j},{\MNC{U}{\ml}{k}}^*\}$, is a solution of Problem~\ref{pro:GIA} almost surely.
\end{Cor}
\begin{IEEEproof}
Please refer to Appendix~\ref{pf_thm:opt} for the proof.
\end{IEEEproof}
\begin{Remark}[Execute Algorithm~\ref{alg:IA} Distributively] Similar to the classical iterative IA algorithm \cite{GomCadJaf:08,GomCadJaf:11}, Algorithm~\ref{alg:IA} can be executed distributively.
To achieve this, after Step 2, LR $k$ needs to send the updated $\tilde{\mathbf{U}}_{k}$ to LTs (or LJs) with index $j:(k,j)\in \mathcal{A}$, and after Step 3,
LT (or LJ) $j$ needs to
send the updated $\tilde{\mathbf{V}}_{j}$ to LRs with index $k:(k,j)\in \mathcal{A
}$.~\hfill~\IEEEQED
\end{Remark}

\section{Numerical Results}
\label{sec:sim}
In this section, we will numerically test the convergence properties of the proposed algorithm, i.e., Algorithm~\ref{alg:IA} and the classical iterative IA algorithm proposed in \cite{GomCadJaf:08}. Please refer to Part II for the numerical results on how GIA techniques enhance secrecy protection.

Consider classical interference networks, i.e., networks with $\tilde{K}=K$, $\mathcal{A}=\mathcal{A}_{\mathrm{all}}$.
To verify if the IA algorithms can always find a solution in IA feasible scenarios,  the following test is adopted.
\begin{Test}[Convergence Test on Random Interference Networks] \label{test} Randomly select configuration within the set\footnote{The sizes
of the networks are restricted so as to maintain manageable computation load.}
\begin{eqnarray}
\nonumber K\in\{3,4,5\},\quad d_k\in{1,2,3},\quad d_k\le M_k,\mnc{N}{\ml}{k} \le
15,\quad \forall k,
\end{eqnarray}
then randomly generate channel state $\{\MNC{H}{\ml}{kj}\}$ following independent complex Gaussian distribution.
First check if the network is IA feasible by testing full row-rankness of matrix $\mathbf{H}_{\mathrm{all}}$
(defined in Fig.~\ref{fig_Hall}). If the network is IA feasible, perform the algorithm to be tested on this network. Denote the output transceivers after $t$ rounds of iteration by $\{\mathbf{V}_k(t),\MNC{U}{\ml}{k}(t)\}$. $\{\mathbf{V}_k(0),\MNC{U}{\ml}{k}(0)\}$ are the initial guesses of the transceivers. Define the normalized power of interference (dB) after $t$ rounds of iteration as
\begin{eqnarray}I(t)=10\log_{10}\frac{\sum_{k=1}^K\sum_{j=1\atop j\neq k}^K
\big|\big|(\MNC{U}{\ml}{k}(t))^\mathrm{H}\MNC{H}{\ml}{kj}\mathbf{V}_{j}(t)\big|\big|^2_\mathrm{F}}
{\sum_{k=1}^K\sum_{j=1\atop j\neq k}^K
\big|\big|(\MNC{U}{\ml}{k}(0))^\mathrm{H}\MNC{H}{\ml}{kj}\mathbf{V}_{j}(0)\big|\big|^2_\mathrm{F}}.\label{eqn:normalIpower}
\end{eqnarray}
If the normalized power of
interference can be reduced  below $-60$ dB after some $t$, the algorithm passes the test. Otherwise, if the algorithm converges to a point with $I(t)>-60$, it fails the test.   ~\hfill~\IEEEQED
\end{Test}

Test~\ref{test} was performed for $10^6$ times on both Algorithm~\ref{alg:IA} and the classical iterative IA algorithm. In all the IA feasible scenarios (about $6.6\times10^5$ cases), both algorithms pass the test. This result verifies the claim of Corollary~\ref{thm:opt}.

To demonstrate how network configuration affects the convergence properties of the proposed algorithm and classical IA algorithm, consider three similar networks
\begin{itemize}
\item{\bf Configuration~1} (Feasible Symmetric Network){\bf:} $\chi=\{(6,6,6),\;(6,6,6),\;(3,3,3)\}$;
\item{\bf Configuration~2} (Feasible Asymmetric Network){\bf:} $\chi=\{(5,5,5),\;(6,6,9),\;(3,3,3)\}$;
\item{\bf Configuration~3} (Infeasible Network){\bf:} $\chi=\{(5,5,5),\;(5,7,9),\;(3,3,3)\}$.
\end{itemize}

\begin{figure}[t] \centering
\includegraphics[scale=0.6]{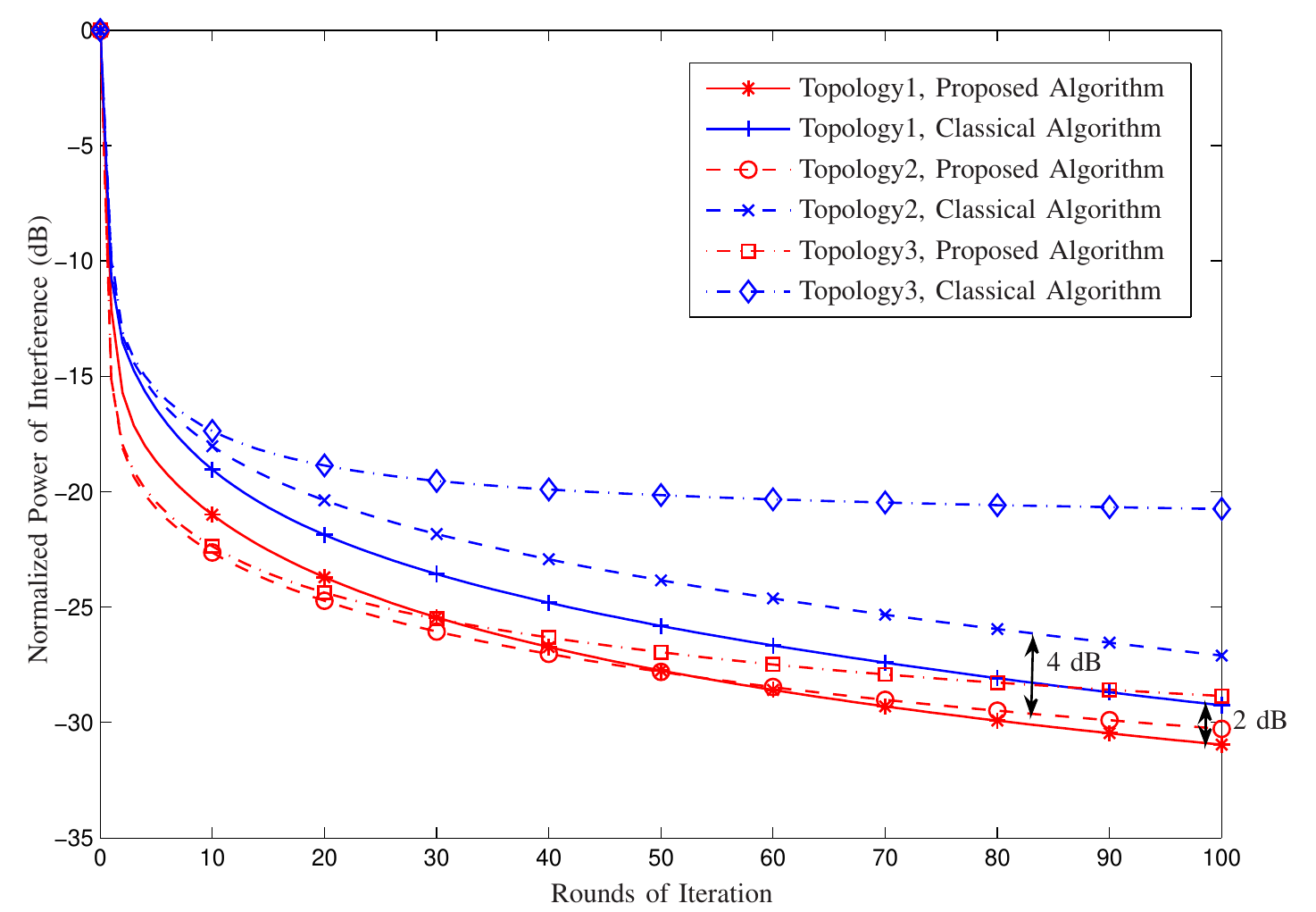}
\caption {Normalized power of interference as a function of rounds of iteration in different network configuration. For fair comparison, we have scaled the output transceivers of both algorithms so that $\sum_{k=1}^K\mathrm{trace}\big(\mathbf{V}^\mathrm{H}_k\mathbf{V}_k\big)$
and $\sum_{k=1}^K\mathrm{trace}\big((\MNC{U}{\ml}{k})^\mathrm{H}\MNC{U}{\ml}{k})$ remain constants.}
\label{fig_convergence}
\end{figure}

Fig.~\ref{fig_convergence} illustrates the normalized power of interference $I(t)$ as a function of rounds of iteration $t$ under the proposed and classical IA algorithms in the three network configurations.
In the two IA feasible networks, both algorithms converges sub-linearly, with the proposed algorithm converging $2$dB and $4$dB faster in the symmetric and asymmetric cases respectively.
In the IA infeasible network, under the classical IA algorithm, $I(t)$ converges to $-21$dB, whereas the proposed algorithm reduces $I(t)$ to $-28$dB after 100 rounds of iteration (and converges to $-30$dB after
400 rounds of iteration).

\section{Summary}
\label{sec:conclude}
In Part I, we have proposed a GIA approach to further improve the IA's
capability in secrecy enhancement.
As illustrated in Fig.~\ref{fig_Framework}, we have established an algebraic
framework
that reveals the (almost sure) equivalence of 1) feasibility of GIA,
2) algebraic independence of GIA constraints, 3) linear independence of the
coefficient vectors of the first order terms in GIA constraints, and 4) full
rankness of the Jacobian matrix of GIA constraints.
This framework allows us to address the two fundamental issues of GIA, i.e.,
feasibility conditions and transceiver design and hence sets up a foundation
for the development and implementation
of GIA (and IA, as a special case) techniques.
\appendices
\section{Proof of Theorem~\ref{thm:F2AI}}
\label{pf_thm:F2AI}
To prove the ``if" side, first prove the following lemma.
\begin{Lem}[Algebraic Independence Leads to Solutions]\label{lem:algdep_sol} $\{c_1,\ldots,c_L\}\in\mathbb{C}^L$ are independent random variables drawn from continuous distribution. Then if polynomials $f_l\in \mathbb{C}(x_1,x_2,\ldots,x_{S})$, $l\in\{1,\ldots,L\}$ are algebraically independent, equation set $f_l=c_l$, $l\in\{1,\ldots,L\}$ has solutions almost surely. Otherwise, the equation set has no solution almost surely.
\end{Lem}
\begin{IEEEproof}
The first half of the lemma is proved in \cite[Lem. 3.2]{RuaLauWin:J13}.
Hence, the focus is on the second half of the lemma.

Denote $F: \mathbb{C}^S\rightarrow \mathbb{C}^L$ as the polynomial map defined by $\{f_l\}$. Since $f_l$ are algebraically dependent, there exists a non-zero polynomial $g$ such that $g(f_1,\ldots,f_L)\equiv0$. Then for any point $[\tilde{c}_1,\ldots,\tilde{c}_L]\in F(\mathbb{C}^S)$, $g(\tilde{c}_1,\ldots,\tilde{c}_L)=0$. On the other hand, since $g$ is a non-zero polynomial, and $\{c_1,\ldots,c_L\}\in\mathbb{C}^L$ are independent random variables drawn from continuous distribution, $g({c}_1,\ldots,{c}_L)\neq 0$ almost surely. Hence, $[{c}_1,\ldots,{c}_L]\not\in F(\mathbb{C}^S)$ almost surely. \end{IEEEproof}

Now turn to the main flow of the proof of the ``if" side. From Lemma~\ref{lem:algdep_sol}, when $\{f_{kjpq}\}$ are algebraically independent, Problem~\ref{pro:IA_poly} has solutions almost surely. Then from \eqref{eqn:czero_poly}, the solution $\{\MNC{U}{\ml}{k},\mathbf{V}_j\}$ constructed by \eqref{eqn:tildeuv} satisfies \eqref{eqn:czero}. Further noting that
\begin{itemize}
\item $\{{\mathbf{U}}_k,{\mathbf{V}}_j\}$ are functions of the channel state of the cross links $\{\MNC{H}{\ml}{kj},k\neq j\}$, and are hence independent of the channel state of the direct links $\{\MNC{H}{\ml}{kk}\}$, and
\item $\mathrm{rank}(\MNC{U}{\ml}{k})=\mathrm{rank}(\mathbf{V}_k)=d_k$,
\end{itemize}
we have that $\{{\mathbf{U}}_k,{\mathbf{V}}_j\}$ constructed by \eqref{eqn:tildeuv} satisfy \eqref{eqn:drank} and \eqref{eqn:rankJ} almost surely. Hence, in this case, Problem~\ref{pro:GIA} has solutions almost surely.

The ``only if" side will be proved by verifying its converse-negative proposition:
\begin{Prop}When $\{f_{kjpq}\}$ are algebraically dependent, Problem~\ref{pro:GIA} has no solution almost surely.\label{prop:AD2NoSol}
\end{Prop}

To prove this proposition, first prove following lemmas.

\begin{Lem}[Algebraic Independence of Random Polynomials]\label{lem:algdepence} The coefficients of polynomials $f_l\in \mathbb{C}(x_1,x_2,\ldots,x_{S})$, $l\in\{1,\ldots,L\}$ are random variables drawn from continuous distribution. Then polynomials $\{f_l\}$ are either always algebraically dependent or algebraically independent almost surely.
\end{Lem}
\begin{IEEEproof}
$\{f_l\}$ are algebraically dependent iff. there exists a non-zero polynomial $g\in\mathbb{C}(y_1,y_2,$ $\ldots, y_{L})$ such that
\begin{eqnarray}g(f_1,\ldots,f_L)\equiv0\label{eqn:depdendent}.
\end{eqnarray}
Without loss of generality, suppose $g$ has $N$ terms, whose coefficients are given by $\{c_1,\ldots,c_N\}$; then \eqref{eqn:depdendent} can be rewritten as a set of linear equations:
\begin{eqnarray}
\mathbf{F}\mathbf{c}=\mathbf{0}.\label{eqn:depdendent_2}
\end{eqnarray}
where $\mathbf{c}=[c_1,\ldots,c_N]^{\mathrm T}$,
$\mathbf{F}\in\mathbb{C}^{S\times N}$, $S$ is the number of terms in $g(f_1,\ldots,f_L)$ after combining like terms. For instance, suppose $f_1=a_1+b_1x_1$, $f_2=a_2+b_2x_2$ and $g=c_1y_1+c_2y_2+c_3y_1y_2$; then
\begin{eqnarray}
g(f_1,f_2)\!\!&=&\!\![a_1,a_2,a_1a_2]\mathbf{c} + [b_1,0,b_1a_2]\mathbf{c}x_1 +\nonumber\\
&&\!\![0,b_2,a_1b_2]\mathbf{c}x_2 + [0,0,b_1b_2]\mathbf{c}x_1x_2.
\end{eqnarray}
Hence, \eqref{eqn:depdendent_2} is given by
\begin{eqnarray}
\begin{bmatrix}
a_1 &a_2 & a_1a_2\\
b_1 &0   & b_1a_2\\
0   &b_2 & a_1b_2\\
0   &0   & b_1b_2\\
\end{bmatrix}
\begin{bmatrix}
c_1\\
c_2\\
c_3
\end{bmatrix}=
\begin{bmatrix}
0\\
0\\
0\\
0
\end{bmatrix}.
\end{eqnarray}

Note that \eqref{eqn:depdendent_2} has non-zero solutions iff. $\mathcal{N}(\mathbf{F})\neq\{0\}$, i.e., $\mathbf{F}$ is column-rank deficient. From Lemma~\ref{lem:rank}, Eq. \eqref{eqn:depdendent_2} either always has no non-zero solutions or has non-zero solutions almost surely. This completes the proof.
\end{IEEEproof}
\begin{Lem}[Rank of a Random Matrix]\label{lem:rank}
Suppose the entries of a matrix $\mathbf{F}\in\mathbb{C}^{M\times N}$ are either 0 or random variables drawn from continuous distribution. Then $\mathbf{F}$ is
 either always column-rank deficient or full column-rank almost surely.
\end{Lem}
\begin{IEEEproof}
If $M<N$, $\mathbf{F}$ is always column-rank deficient. Otherwise, denote all the $N\times N$ submatrices in $\mathbf{F}$ by $\{\tilde{\mathbf{F}}_{1},\ldots,\tilde{\mathbf{F}}_{D}\}$, where $D=\begin{pmatrix}M\\N\end{pmatrix}$; then $\mathbf{F}$ is full column-rank iff. the determinant of at least one $\tilde{\mathbf{F}}_{d}$, $d\in\{1,\ldots,D\}$ is not zero.

From the Leibniz formula \cite[6.1.1]{Mey:00}, the determinant $\tilde{\mathbf{F}}_{d}$ is given by a polynomial of the entries in $\tilde{\mathbf{F}}_{d}$. If this polynomial is a zero polynomial, the determinant of $\tilde{\mathbf{F}}_{d}$ is always 0. Otherwise, noting that the entries of $\tilde{\mathbf{F}}_{d}$ are drawn from continuous distribution, the value of this polynomial is non-zero almost surely.  This completes the proof.
\end{IEEEproof}

Now turn to the main flow of the proof of Proposition~\ref{prop:AD2NoSol}.
Consider a solution $\{{\MNC{U}{\ml}{k}}^*,\mathbf{V}^*_j\}$ of Problem~\ref{pro:GIA}.
From \eqref{eqn:drank}, we have that
$\mathrm{rank}({\MNC{U}{\ml}{k}}^*)=d_k$ and $\mathrm{rank}(\mathbf{V}^*_j)=d_j$, $\forall k,j$. Hence, every ${\MNC{U}{\ml}{k}}^*$ (or $\mathbf{V}^*_j$) has at least $d_k$ (or $d_j$) linearly independent row vectors. Denote the submatrices aggregated by these linearly independent rows by $\mathbf{U}^{(1)}_k$ (or $\mathbf{V}^{(1)}_k$). Transform $\mathbf{U}_{k},\mathbf{V}_{j}$ as follows:
\begin{eqnarray}
\mathbf{U}'_k
\!=\!\MNC{U}{\ml}{k}\Big(\mathbf{U}^{(1)}_k\Big)^{\!-1}\!\!\!,\; \mathbf{V}'_j\!=\!\mathbf{V}_j\Big(\mathbf{V}^{(1)}_j\Big)^{\!-1}\!\!\!,
\label{eqn:tildeuv2}
\end{eqnarray}
and let $\tilde{\mathbf{U}}_{k}$ and $\tilde{\mathbf{V}}_{j}$ be the nonconstant parts in $\mathbf{U}'_{k}$ and $\mathbf{V}'_{j}$, respectively.
Then, $\{\tilde{\mathbf{U}}_{k},\tilde{\mathbf{V}}_{j}\}$ satisfies a set of polynomial equations in the same form as \eqref{eqn:czero_poly}, in which the position of $\{\mathbf{U}^{(1)}_k,\mathbf{V}^{(1)}_k\}$ in $\{{\mathbf{U}}'_{k},{\mathbf{V}}'_{j}\}$ only affects the indices of the coefficients. For example, suppose $\mathbf{U}^{(1)}_k$, $\mathbf{V}^{(1)}_j$ are given by the last $d_k\times d_k$ and $d_j\times d_j$ submatrices in $\MNC{U}{\ml}{k}$ and $\mathbf{V}_j$ respectively; then \eqref{eqn:czero} can be rewritten as
\begin{eqnarray}
\nonumber f_{kjpq}\!\!&\triangleq&\!\! \tilde{\mathbf{u}}^\mathrm{H}_{k}(p)\MNC{H}{\ml}{kj}(1:\!\mnc{N}{\ml}{k}\!-\!d_k,M_j\!-\!d_j\!+\!q)
\\\nonumber\!\!&&\!\!+\MNC{H}{\ml}{kj}(\mnc{N}{\ml}{k}\!-\!d_k\!+\!p,1:\!M_j\!-\!d_j)\tilde{\mathbf{v}}_{j}(q)
\\\nonumber \!\!&&\!\!+\tilde{\mathbf{u}}^\mathrm{H}_{k}(p)
\MNC{H}{\ml}{kj}(1:\mnc{N}{\ml}{k}\!-\!d_k,1:M_j\!-\!d_j)\tilde{\mathbf{v}}_{j}(q)
\\\!\! &=&\!\! -h_{kj}(\mnc{N}{\ml}{k}\!-\!d_k\!+\!p,M_j\!-\!d_j\!+\!q), \label{eqn:czero_poly2}
\end{eqnarray}
which is the same as \eqref{eqn:czero_poly}, except for the indices of the coefficients.

Since all entries the of channel state matrices $\MNC{H}{\ml}{kj}$ are independent random variables drawn from continuous distribution, we have that if Problem~\ref{pro:IA_poly} has no solution almost surely, for every possible position of $\{\mathbf{U}^{(1)}_k,\mathbf{V}^{(1)}_k\}$, the corresponding equation set also has no solution almost surely. Hence, Problem~\ref{pro:GIA} has no solution almost surely.

\section{Proof of Theorem~\ref{thm:AI2LI}}
\label{pf_thm:AI2LI}
The proof of the first statement in Theorem~\ref{thm:AI2LI} is given by Lemma~\ref{lem:rank}.

If matrix $\mathbf{H}_{\mathrm{all}}$ is full row-rank almost surely, from \cite[Lem. 3.1]{RuaLauWin:J13}, polynomials $\{f_{kjpq}\}$ are algebraically independent almost surely. Hence, the focus is on the other case.

The size of matrix $\mathbf{H}_{\mathrm{all}}$ is $C\times V$, where $C=\sum_{k=1}^{K}\!\sum_{j=1,\atop(k,j)\in\mathcal{A}}^{\tilde{K}} d_kd_j$ and $ V=\sum_{k=1}^{K}d_k(\mnc{N}{\ml}{k}-d_k)+\sum_{j=1}^{\tilde{K}}d_j(M_j-d_j)$. If matrix $\mathbf{H}_{\mathrm{all}}$ is always row-rank deficient, there are two possibilities:
\begin{itemize}
\item{\emph When $C>V$:} Denote $\mathcal{V}$ as the set of all entries in $\tilde{\mathbf{U}}_k$ and $\tilde{\mathbf{V}}_j$, $k\in\{1,\ldots,K\}$, $j\in\{1,\ldots,\tilde{K}\}$. From \cite[Cor. 5.7]{Kem:10}, the dimension of the field $\mathcal{V}]$ is $V$. On the other hand, the number of the polynomials in $\{f_{kjpq}\}$, i.e., $C$, is greater than $V$. Hence, from \cite[Def. 5.3]{Kem:10}, $\{f_{kjpq}\}$ must be algebraically dependent.
\item{\emph When $C\le V$:} Denote all the $C\times C$ submatrices in $\mathbf{H}_{\mathrm{all}}$ by $\{\tilde{\mathbf{H}}_{1},\ldots,\tilde{\mathbf{H}}_{D}\}$, where $D=\begin{pmatrix}V\\C\end{pmatrix}$. Since $\mathbf{H}_{\mathrm{all}}$ is always row-rank deficient,
    \begin{eqnarray}\det(\tilde{\mathbf{H}}_{d})\equiv 0\label{eqn:detzero}
    \end{eqnarray}
    for all $d\in\{1,\ldots,D\}$ and all possible channel states $\{\MNC{H}{\ml}{kj}\}$. From the Leibniz formula, $\det(\tilde{\mathbf{H}}_{d})$ is given by a polynomial of the entries in $\tilde{\mathbf{H}}_{d}$. Denote this polynomial by $g_d(\{h_{kj}(p,q)\})\triangleq \det(\tilde{\mathbf{H}}_{d})$, $k\in\{1,\ldots, K\}$, $j\in\{1,\ldots, \tilde{K}\}$, $(p,q)\in \big(\{d_k+1,\ldots,\mnc{N}{\ml}{k}\}\times\{1,\ldots,d_j\}\big)\cup\big(\{1,\ldots,d_k\}\times \{d_j+1,\ldots,M_j\}\big)$. Then from \eqref{eqn:detzero}, $g_d$ are zero polynomials for all $d\in\{1,\ldots,D\}$.

Next, consider the Jacobian matrix of $\{f_{kjpq}\}$, i.e.,  $\mathbf{J}_{\tilde{\mathbf{U}}_k,\tilde{\mathbf{V}}_j}(\{f_{kjpq}\})$. From \eqref{eqn:czero_poly}, $\mathbf{J}_{\{\tilde{\mathbf{U}}_k,\tilde{\mathbf{V}}_j\}}(\{f_{kjpq}\})$ has the same structure as $\mathbf{H}_{\mathrm{all}}$, with the following differences:
    \begin{eqnarray}
    \hspace{-2mm}\left\{\begin{array}{l@{\,}l}
    \mbox{In \eqref{eqn:hu}, }&h_{kj}(p,q), p\in\{d_k\!+\!1,\ldots,\mnc{N}{\ml}{k}\},\\     &q\in\{1,\ldots,d_j\}\mbox{ is replaced by }
    \\&h_{kj}(p,q)+\MNC{H}{\ml}{kj}(p,d_j\!+\!1:M_j)\tilde{\mathbf{v}}_{j}(q);
    \\\mbox{In \eqref{eqn:hv}, }&h_{kj}(p,q), p\in\{1,\ldots,d_k\},\\
    &q\in\{d_j\!+\!1,\ldots,M_j\} \mbox{ is replaced by }
    \\&h_{kj}(p,q)+\tilde{\mathbf{u}}^\mathrm{H}_{k}(p)
    \MNC{H}{\ml}{kj}(d_k\!+\!1:\mnc{N}{\ml}{k},q).
    \end{array}\right.\label{eqn:huv_change}
    \end{eqnarray}
Denote all the $C\times C$ submatrices in $\mathbf{J}_{\{\tilde{\mathbf{U}}_k,\tilde{\mathbf{V}}_j\}}(\{f_{kjpq}\})$ by $\{\tilde{\mathbf{J}}_{1},\ldots,\tilde{\mathbf{J}}_{D}\}$. Define linear functions $\ell_{kjpq}(\tilde{\mathbf{U}}_k,\tilde{\mathbf{V}}_j)$ as
\begin{eqnarray}
    &&\hspace{-10mm}\ell_{kjpq}(\tilde{\mathbf{U}}_k,\tilde{\mathbf{V}}_j)=\nonumber
    \\&&\hspace{-5mm}\left\{\begin{array}{l}
    h_{kj}(p,q)+\MNC{H}{\ml}{kj}(p,d_j\!+\!1:M_j)\tilde{\mathbf{v}}_{j}(q),
    \\\mbox{if: } p\in\{d_k\!+\!1,\ldots,\mnc{N}{\ml}{k}\}, q\in\{1,\ldots,d_j\};
    \\h_{kj}(p,q)+\tilde{\mathbf{u}}^\mathrm{H}_{k}(p)
    \MNC{H}{\ml}{kj}(d_k\!+\!1:\mnc{N}{\ml}{k},q),
    \\\mbox{if: } p\in\{1,\ldots,d_k\}, q\in \{d_j\!+\!1,\ldots,M_j\}.\end{array}\right.\label{eqn:Lform}
\end{eqnarray}

Then from \eqref{eqn:huv_change}, noticing the one to one correspondence between $\{h_{kj}(p,q)\}$ and $\{\ell_{kjpq}(\tilde{\mathbf{U}}_k,\tilde{\mathbf{V}}_j)\}$,
$\det\big(\mathbf{J}_{d}\big)$ can be written as the cascade of $g_d$ and $\{\ell_{kjpq}\}$, i.e.,
    \begin{eqnarray}\det\big(\mathbf{J}_{d}\big)
    =g_d(\{\ell_{kjpq}(\tilde{\mathbf{U}}_k,\tilde{\mathbf{V}}_j)\}),\quad d\in\{1,\ldots,D\},\label{eqn:Jsubdet}\end{eqnarray}

    Since $\{g_d\}$ are zero polynomials, $\det\big(\mathbf{J}_{d}\big)\equiv0$, $\forall d\in\{1,\ldots,D\}$, which means that $\mathbf{J}_{\{\tilde{\mathbf{U}}_k,\tilde{\mathbf{V}}_j\}}(\{f_{kjpq}\})$ is always row-rank deficient.  From \cite[Thm. 2.3]{EhrRot:93}, $\{f_{kjpq}\}$ are algebraically dependent.
\end{itemize}

\section{Proof of Theorem~\ref{thm:AI2J}}
\label{pf_thm:AI2J}

From \cite[Thm. 2.2]{EhrRot:93}, when $\mathbf{J}_{\mathbf{x}}(\{f_{kjpq}\})$, is not always row-rank deficient, $\{f_{kjpq}\}$ are algebraically independent. Hence, the ``if" side is proved.

The ``only if" side is true if the following lemma holds:
\begin{Lem}\label{lem:AIJrank}
If $\{f_{kjpq}\}$ are algebraically independent, $\mathbf{J}_{\mathbf{x}}(\{f_{kjpq}\})$ is row-rank deficient on a proper closed subset of $\mathbb{C}^V$.
\end{Lem}

From \eqref{eqn:Jsubdet} and \eqref{eqn:Lform}, the set in which $\mathbf{J}_{\mathbf{x}}(\{f_{kjpq}\})$ is row-rank deficient is given by $\cap_{d=1}^D\mathcal{N}_d$, where
\begin{eqnarray}
\mathcal{N}_d\triangleq \big\{\tilde{\mathbf{U}}_k,\tilde{\mathbf{V}}_j,k\in\{1,\ldots,K\},j\in\{1,\ldots,\tilde{K}\}:\nonumber\\
g_d(\{\ell_{kjpq}(\tilde{\mathbf{U}}_k,\tilde{\mathbf{V}}_j)\})=0\big\}
\end{eqnarray}

If $g_d(\{\ell_{kjpq}(\tilde{\mathbf{U}}_k,\tilde{\mathbf{V}}_j)\})$ is a zero polynomial of $\{\tilde{\mathbf{U}}_k,\tilde{\mathbf{V}}_j\}$, $\mathcal{N}_d=\mathbb{C}^V$; otherwise, $\mathcal{N}_d$ is a proper closed set of $\mathbb{C}^V$. When $\{f_{kjpq}\}$ are algebraically independent, at least one $g_d(\{\ell_{kjpq}(\tilde{\mathbf{U}}_k,\tilde{\mathbf{V}}_j)\})$ is a non-zero polynomial. Further noting that the intersection of closed sets is closed, Lemma~\ref{lem:AIJrank} is proved.

\section{Proof of Corollary~\ref{cor:sym}}
\label{pf_cor:sym}
\begin{figure}[t] \centering
\includegraphics[scale=0.52]{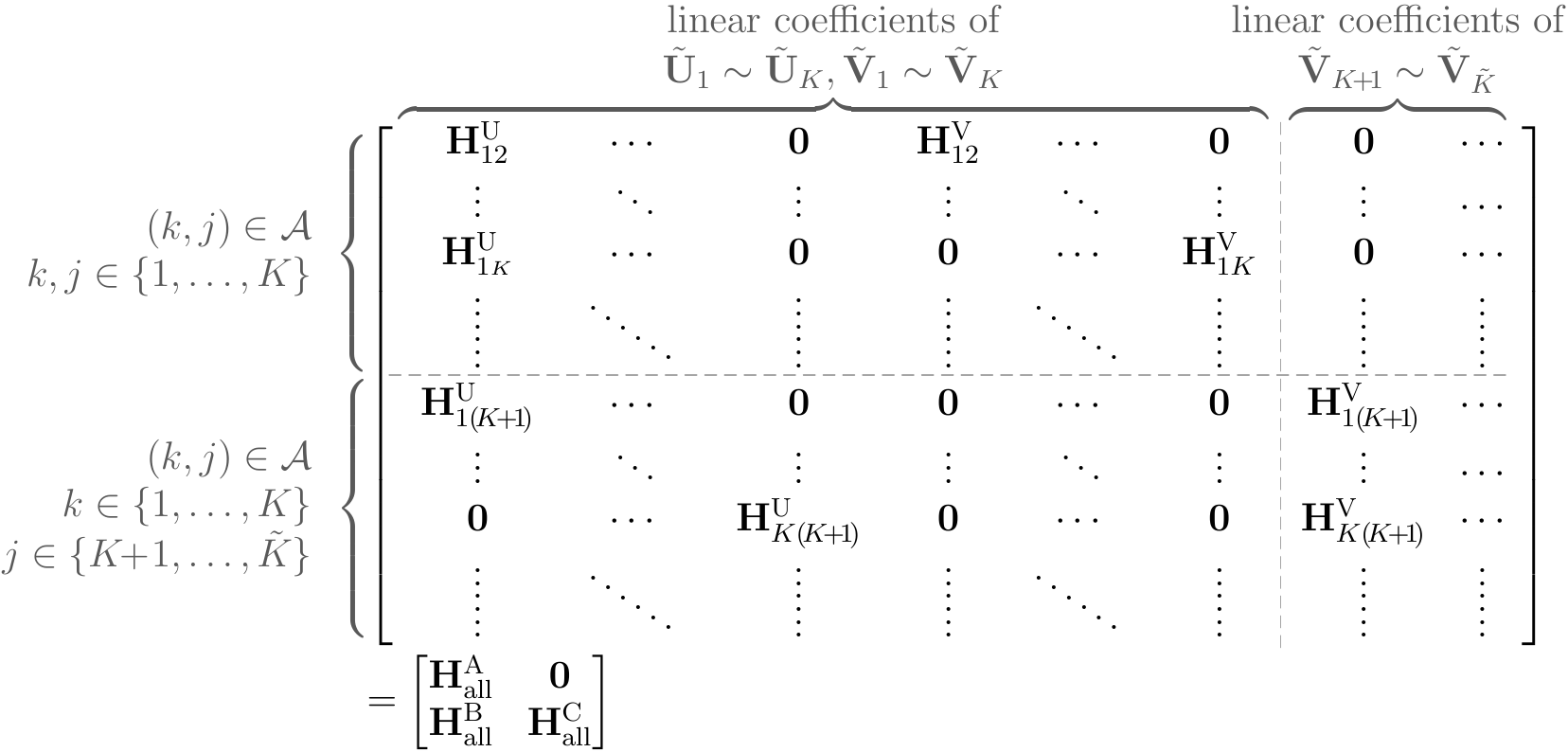}
\caption {Row-switching and separation of $\mathbf{H}_{\mathrm{all}}$. For clear illustration, we have set $\mathcal{A}=\mathcal{A}_{\mathrm{all}}$ when plotting the figure.}
\label{fig_Hall_Cor1}
\end{figure}

From Theorem~\ref{thm:feasible}, one needs to show that $\mathbf{H}_{\mathrm{all}}$ is full row rank iff. \eqref{eqn:f_sym} is true. As illustrated in Fig.~\ref{fig_Hall_Cor1}, perform row switching and then separate $\mathbf{H}_{\mathrm{all}}$ into four submatrices, i.e., $\mathbf{H}^{\mathrm{A}}_{\mathrm{all}}$--$\mathbf{H}^{\mathrm{C}}_{\mathrm{all}}$ and one zero matrix. The following lemma shows the full rankness of $\mathbf{H}^{\mathrm{C}}_{\mathrm{all}}$.

\begin{Lem}[Full rankness of $\mathbf{H}^{\mathrm{C}}_{\mathrm{all}}$]\label{lem:HCrank} Under condition 3) in Corollary~\ref{cor:sym},  $\mathbf{H}^{\mathrm{C}}_{\mathrm{all}}$ is full row rank almost surely.
\end{Lem}
\begin{IEEEproof} Note that \begin{eqnarray}
\mathbf{H}^{\mathrm{C}}_{\mathrm{all}}=\mathrm{diag}\big(\mathbf{H}^{\mathrm{C}}_{K+1},
\ldots,\mathbf{H}^{\mathrm{C}}_{\tilde{K}}\big),\label{eqn:HC}\end{eqnarray}
where $\mathbf{H}^{\mathrm{C}}_{j}$, $j\in\{K+1,\ldots,\tilde{K}\}$ is aggregated by submatrices $\mathbf{H}^{\mathrm{V}}_{kj}$, $\forall k:(k,j)\in\mathcal{A}$. From the structure of $\mathbf{H}^{\mathrm{V}}_{kj}$ in \eqref{eqn:hv}, by doing row switching operations, $\mathbf{H}^{\mathrm{C}}_{j}$ can be transformed into a block diagonal matrix with $d_j$ diagonal blocks. Note that
\begin{itemize}
\item[(a)] the size of these diagonal blocks is $\big(d\sum_{k=1}^K\mathbb{I}\{(k,j)\in\mathcal{A}\}\big)\times \big(M_j-d_j\big)$;
\item[(b)] within each diagonal block, all entries are independent random variables.
\end{itemize}
Hence, when condition 3) in Corollary~\ref{cor:sym} holds, the diagonal blocks in $\mathbf{H}^{\mathrm{C}}_{j}$ are full row-rank almost surely. Therefore, $\mathbf{H}^{\mathrm{C}}_{j}$ is full row-rank almost surely. Substituting this result to \eqref{eqn:HC},   $\mathbf{H}^{\mathrm{C}}_{\mathrm{all}}$ is full row-rank almost surely. This completes the proof.\end{IEEEproof}

With Lemma~\ref{lem:HCrank}, and further noting that $\mathbf{H}_{\mathrm{all}}$ is a block-upper-triangular matrix, the corollary holds if the following proposition is true:
\begin{Prop}\label{prop:HArank} Under condition 1) and 2) in Corollary~\ref{cor:sym},
$\mathbf{H}^{\mathrm{A}}_{\mathrm{all}}$ is full row rank iff. \eqref{eqn:f_sym} is true.
\end{Prop}

When \eqref{eqn:f_sym} is not satisfied, $\mathbf{H}^{\mathrm{A}}_{\mathrm{all}}$
is row-rank deficient as it has more rows than columns.
Hence, the ``only if" statement in Proposition~\ref{prop:HArank} is proved. The ``if" side can be proved via the following steps:
 \begin{itemize}
 \item[A.] Construct one special category of channel state $\{\mathbf{H}_{kj}\}$.
 \item[B.] Show that $\mathbf{H}_{\mathrm{all}}$ is full rank almost surely under the special category of channel state.
 \item[C.] From the first statement in Theorem~\ref{thm:AI2LI}, if Procedure B is completed, $\mathbf{H}^{\mathrm{A}}_{\mathrm{all}}$ is full rank almost surely, and this proves the corollary.
\end{itemize}

\begin{figure}[t] \centering
\includegraphics[scale=0.7]{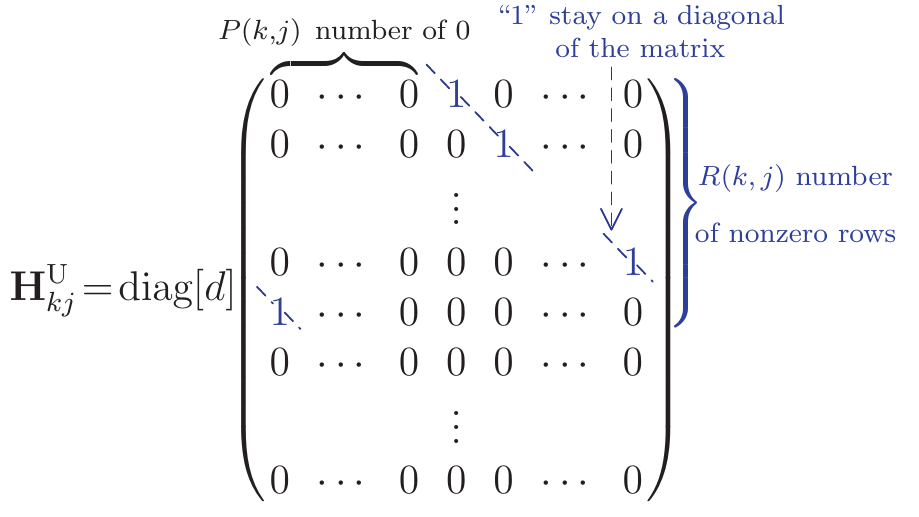}
\caption {Specify $\{\mathbf{H}^{\mathrm{U}}_{kj}\}$.}
\label{fig_specifyhu}
\end{figure}

Construct a special $\mathbf{H}_{\mathrm{all}}$ by using tools from graph theory. Consider a graph $\mathcal{G}$ whose vertexes are the nodes of the network and there is an edge between LT $j$ and LR $k$, if $(k,j)\in\mathcal{A}$. Then from \cite[Thm. 8.15]{HsuLin:B09}, when the alignment set is $L$-regular, there is a proper $L$-edge-coloring \cite[Page 138]{HsuLin:B09} for the graph. Denote the coloring of an edge between LT $j$ and LR $k$ by $f(k,j)\in\{1,2,\ldots,L\}$ and specify $\{\mathbf{H}^{\mathrm{U}}_{kj}\}$ as in Fig.~\ref{fig_specifyhu}, in which
\begin{eqnarray}
\!\!\!\!\!\!\!P(k,j)\!\!&=&\!\!d(f(k,j)\!-\!1) \;\mathrm{mod}\;
(N\!-\!d) \label{eqn:pos}\\
\!\!\!\!\!\!\!R(k,j)\!\!&=&\!\!\left\{\begin{array}{ll}
d &\mbox{if }f(k,j)\!\le\! \lfloor\!\frac{N}{d}\!\rfloor,\\
(N\!-\!1)\;\mathrm{mod}\;d&\mbox{if }f(k,j)\!=\! \lfloor\!\frac{N}{d}\!\rfloor\!+\!1,\\
0 &\mbox{otherwise.}
\end{array}\right.\label{eqn:rownumber}
\end{eqnarray}
The rest of the proof is similar to that of Cor. 3.3 in \cite{RuaLauWin:J13}.

\section{Proof of Corollary~\ref{cor:div}}
\label{pf_cor:div}
The proof is similar to that of \cite[Cor. 3.4]{RuaLauWin:J13}. To accommodate the alignment set $\mathcal{A}$, one need to change equations (24) and (25) in \cite{RuaLauWin:J13} to
\begin{eqnarray}
&&\hspace{-10mm}c^{\mathrm{t}}_{kjpq}+c^{\mathrm{r}}_{kjpq}=\left\{\begin{array}{ll}
1 & \mbox{if: } (k,j)\in\mathcal{A}\\
0 & \mbox{otherwise}
\end{array}\right.,\label{eqn:cv1}
\\&&\hspace{-10mm}\sum_{j=1,\neq k}^{\tilde{K}} \sum_{q=1}^{d_j} c^{\mathrm{r}}_{kjpq}\le \mnc{N}{\ml}{k}-d_k,\quad \forall k\in\{1,\ldots,K\}, \label{eqn:cv2r}
\end{eqnarray}
respectively. Then the rest of the proof follows. The details are omitted to avoid redundancy.

\section{Proof of Theorem~\ref{thm:nogap}}
\label{pf_thm:nogap}
The theorem will be proved by contradiction.

Suppose there exists a local optimum $\{\tilde{\mathbf{U}}^\mathrm{L}_{k},\tilde{\mathbf{V}}^\mathrm{L}_{j}\}$ such that \begin{eqnarray}
F(\{g_{kjpq}(\tilde{\mathbf{U}}^\mathrm{L}_{k},\tilde{\mathbf{V}}^\mathrm{L}_{j})\})
> 0.\label{eqn:nonzero}
\end{eqnarray}

$\mathbf{J}_{\{\tilde{\mathbf{U}}_k,\tilde{\mathbf{V}}_j\}}(\{g_{kjpq}\})
=\mathbf{J}_{\{\tilde{\mathbf{U}}_k,\tilde{\mathbf{V}}_j\}}(\{f_{kjpq}\})$.
Hence, from Theorem~\ref{thm:F2AI} and~\ref{thm:AI2J}, when IA is feasible, the set $\{\{\tilde{\mathbf{U}}_k,\tilde{\mathbf{V}}_j\}:\mathbf{J}_{\{\tilde{\mathbf{U}}_k,\tilde{\mathbf{V}}_j\}}(\{g_{kjpq}\})
\mbox{ is full row rank.}\}$ is dense. Therefore,
for any $\delta>0$, there exists a $\{\tilde{\mathbf{U}}_{k},\tilde{\mathbf{V}}_{j}\}$ satisfying:
\begin{eqnarray}
\hspace{-7mm}&&\sum_{k}||\tilde{\mathbf{U}}^\mathrm{L}_{k}-\tilde{\mathbf{U}}_{k}||_\mathrm{F}+
\sum_{j}||\tilde{\mathbf{V}}^\mathrm{L}_{j}-\tilde{\mathbf{V}}_{j}||_\mathrm{F}\le\delta^2\label{eqn:delta2}\\
\hspace{-7mm}&&\mathbf{J}_{\{\tilde{\mathbf{U}}_k,\tilde{\mathbf{V}}_j\}}(\{g_{kjpq}\})
\mbox{ is full row rank.}\label{eqn:fullrank}
\end{eqnarray}

Since both $F$ and all $g_{kjpq}$ continuously differentiable, $\mathbf{J}_{\{\tilde{\mathbf{U}}_k,\tilde{\mathbf{V}}_j\}}(F)$ is finite on any bounded close set.
Therefore, from \eqref{eqn:delta2}, there exists some finite constant $C\ge 0$ such that
\begin{eqnarray}
\left|F(\{g_{kjpq}(\tilde{\mathbf{U}}^\mathrm{L}_{k},\tilde{\mathbf{V}}^\mathrm{L}
_{j})\})-F(\{g_{kjpq}(\tilde{\mathbf{U}}_{k},\tilde{\mathbf{V}}
_{j})\})\right|\le C\delta^2.\label{eqn:delta3}
\end{eqnarray}

When a matrix $\mathbf{A}$ is full row rank, the linear equation set $\mathbf{A}\mathbf{x}=\mathbf{b}$ has solution for any vector $\mathbf{b}$. Therefore, from \eqref{eqn:fullrank}, there exists $\{\Delta\tilde{\mathbf{U}}_{k},\Delta\tilde{\mathbf{V}}_{j}\}$ that satisfies linear equation set
\begin{eqnarray}
\mathbf{J}_{\{\tilde{\mathbf{U}}_k,\tilde{\mathbf{V}}_j\}}(\{g_{kjpq}\})
\begin{bmatrix}
\mathrm{vec}\{\Delta\tilde{\mathbf{U}}_{1}\}
\\\vdots
\\\mathrm{vec}\{\Delta\tilde{\mathbf{U}}_{K}\}
\\\mathrm{vec}\{\Delta\tilde{\mathbf{V}}_{1}\}
\\\vdots
\\\mathrm{vec}\{\Delta\tilde{\mathbf{V}}_{\tilde{K}}\}
\end{bmatrix}\!\!=\!\!
\begin{bmatrix}
g_{1211}(\tilde{\mathbf{U}}_1,\tilde{\mathbf{V}}_2)
\\\vdots
\\g_{kjpq}(\tilde{\mathbf{U}}_k,\tilde{\mathbf{V}}_j)
\\\vdots
\\g_{K\tilde{K}d_Kd_{\tilde{K}}}(\tilde{\mathbf{U}}_K,\tilde{\mathbf{V}}_{\tilde{K}})
\end{bmatrix}.\label{eqn:direction}
\end{eqnarray}

From \eqref{eqn:direction},

\begin{eqnarray}
\nonumber\hspace{-3mm}&\hspace{-3mm}&\begin{bmatrix}
g_{1211}(\tilde{\mathbf{U}}_{1}\!-
\!\delta\Delta\tilde{\mathbf{U}}_{1},
\tilde{\mathbf{V}}_{2}\!-
\!\delta\Delta\tilde{\mathbf{V}}_{2})
\\\vdots
\\g_{K\tilde{K}d_Kd_{\tilde{K}}}(\tilde{\mathbf{U}}_{K}\!-
\!\delta\Delta\tilde{\mathbf{U}}_{K},
\tilde{\mathbf{V}}_{\tilde{K}}\!-
\!\delta\Delta\tilde{\mathbf{V}}_{\tilde{K}})
\end{bmatrix}
\\\nonumber\hspace{-3mm}&=\hspace{-3mm}&
\begin{bmatrix}
g_{1211}(\tilde{\mathbf{U}}_1,\tilde{\mathbf{V}}_2)
\\\vdots
\\g_{K\tilde{K}d_Kd_{\tilde{K}}}(\tilde{\mathbf{U}}_K,\tilde{\mathbf{V}}_{\tilde{K}})
\end{bmatrix}-\nonumber
\\&&\delta\mathbf{J}_{\{\tilde{\mathbf{U}}_k,\tilde{\mathbf{V}}_j\}}(\{f_{kjpq}\})
\begin{bmatrix}
\mathrm{vec}\{\Delta\tilde{\mathbf{U}}_{1}\}
\\\vdots
\\\mathrm{vec}\{\Delta\tilde{\mathbf{V}}_{K}\}
\end{bmatrix}+\Delta\mathbf{ g}
\\\hspace{-3mm}&=&(1-\delta)\begin{bmatrix}
g_{1211}(\tilde{\mathbf{U}}_1,\tilde{\mathbf{V}}_2)
\\\vdots
\\g_{K\tilde{K}d_Kd_{\tilde{K}}}(\tilde{\mathbf{U}}_K,\tilde{\mathbf{V}}_{\tilde{K}})
\end{bmatrix}+\Delta\mathbf{ g},\label{eqn:delta}
\end{eqnarray}
where $||\Delta\mathbf{g}||\sim\mathcal{O}(\delta^2)$. Denote $\Delta\mathbf{g}$ by $\big[\Delta g_{1211},\ldots,\Delta g_{kjpq},\ldots, \Delta g_{K\tilde{K}d_Kd_{\tilde{K}}}\big]^\mathrm{T}$.

Further note that $F$ is convex, continuously differentiable and $F(0,\ldots,0)=0$.
From \eqref{eqn:delta}, there exits some constant $\tilde{C}>0$ such that
\begin{eqnarray}
\nonumber &&\hspace{-10mm}F(\{g_{kjpq}(\tilde{\mathbf{U}}_{k}\!-
\!\delta\Delta\tilde{\mathbf{U}}_{k},
\tilde{\mathbf{V}}_{j}\!-
\!\delta\Delta\tilde{\mathbf{V}}_{j})\})
\\&=&
F(\{(1-\delta)g_{kjpq}(\tilde{\mathbf{U}}_{k},
\tilde{\mathbf{V}}_{j}) + \Delta g_{kjpq}\})
\nonumber
\\&\le& F(\{(1-\delta)g_{kjpq}(\tilde{\mathbf{U}}_{k},
\tilde{\mathbf{V}}_{j})\}) + \tilde{C}\delta^2
\nonumber
\\&\le& \delta F(0,\ldots,0) + (1-\delta) F(\{g_{kjpq}(\tilde{\mathbf{U}}_{k},
\tilde{\mathbf{V}}_{j})\})+ \tilde{C}\delta^2
\nonumber
\\&=&(1-\delta)F(\{g_{kjpq}(\tilde{\mathbf{U}}_{k},
\tilde{\mathbf{V}}_{j})\})+\tilde{C}\delta^2.\label{eqn:delta4}
\end{eqnarray}

From \eqref{eqn:delta3} and \eqref{eqn:delta4},
\begin{eqnarray}
\nonumber&&\hspace{-10mm}F(\{g_{kjpq}(\tilde{\mathbf{U}}^\mathrm{L}_{k},
\tilde{\mathbf{V}}^\mathrm{L}_{j})\})
\!-\!F(\{g_{kjpq}(\tilde{\mathbf{U}}_{k}\!-
\!\delta\Delta\tilde{\mathbf{U}}_{k},
\tilde{\mathbf{V}}_{j}\!-
\!\delta\Delta\tilde{\mathbf{V}}_{j})\})
\\\nonumber&=&\hspace{-3mm}\Big(F(\{g_{kjpq}(\tilde{\mathbf{U}}^\mathrm{L}_{k},
\tilde{\mathbf{V}}^\mathrm{L}_{j})\})
-F(\{g_{kjpq}(\tilde{\mathbf{U}}_{k},
\tilde{\mathbf{V}}_{j})\})\Big)+
\\\nonumber&&\hspace{-3mm}\Big(F(\{g_{kjpq}(\tilde{\mathbf{U}}_{k},
\tilde{\mathbf{V}}_{j})\})-
\\\nonumber&&\hspace{-1.2mm}F(\{g_{kjpq}(\tilde{\mathbf{U}}_{k}\!-
\!\delta\Delta\tilde{\mathbf{U}}_{k},
\tilde{\mathbf{V}}_{j}\!-
\!\delta\Delta\tilde{\mathbf{V}}_{j})\})\Big)
\\\nonumber &\ge&\hspace{-3mm} -C\delta^2+ \delta F(\{g_{kjpq}(\tilde{\mathbf{U}}_{k},\tilde{\mathbf{V}}_{j})\}) -
\tilde{C}\delta^2
\\&\ge&\hspace{-3mm}
\delta F(\{g_{kjpq}(\tilde{\mathbf{U}}^\mathrm{L}_{k},\tilde{\mathbf{V}}^{\mathrm{L}}_{j})\})-
(C+\tilde{C})\delta^2 - C\delta^3.
\label{eqn:difference}
\end{eqnarray}

If \eqref{eqn:nonzero} is true, when $\delta$ is sufficiently small, \eqref{eqn:difference} is positive, which contradicts the assumption that $\{\tilde{\mathbf{U}}^\mathrm{L}_{k},\tilde{\mathbf{V}}^\mathrm{L}_{j}\}$ is a local optimum. This completes this proof.

 \section{Proof of Corollary~\ref{thm:opt}}
\label{pf_thm:opt}
Function  $F(\{x_i\}) =\sum_{i}x_ix_i^\mathrm{H}$ is convex and continuously differentiable, with $F(\{0\})=0$. Hence, from Theorem~\ref{thm:nogap}, one only needs to show that
the output of Algorithm~\ref{alg:IA}, i.e., $\{\mathbf{V}^{*}_{j},{\MNC{U}{\ml}{k}}^*\}$,
is a local optimum.

In Step 2 and 3 of Algorithm~\ref{alg:IA},
the updated $\tilde{\mathbf{U}}_k$, and $\tilde{\mathbf{V}}_j$, given by \eqref{eqn:alg_U} and \eqref{eqn:alg_V} are respectively the optimal solutions of the following two sets of unconstraint quadratic optimization problems:
\begin{Prob}[Interference Optimization at LR $k$]
\begin{eqnarray}
\hspace{-3mm}&\underset{\tilde{\mathbf{U}}_k}{\mbox{minimize}}&
\sum_{j:(j,k)\in\mathcal{A}}\bigg|\bigg|\begin{bmatrix}\mathbf{I}_{d_k\times
d_k}\\ \tilde{\mathbf{U}}_{k}\end{bmatrix}^\mathrm{H}\mathbf{H}_{kj}\mathbf{V}_j\bigg|\bigg|^2_{\mathrm F}\label{eqn:leakage_3}
\end{eqnarray}\end{Prob}

\begin{Prob}[Interference Optimization at LT $j$]
\begin{eqnarray}
\hspace{-3mm}&\underset{\tilde{\mathbf{V}}_j}{\mbox{minimize}}&
\sum_{k:(j,k)\in\mathcal{A}}\bigg|\bigg|\mathbf{U}^\mathrm{H}_k\mathbf{H}_{kj}\begin{bmatrix}\mathbf{I}_{d_j\times
d_j}\\ \tilde{\mathbf{V}}_{j}\end{bmatrix}\bigg|\bigg|^2_{\mathrm F}\label{eqn:leakage_4}
\end{eqnarray}\end{Prob}

Therefore, $\sum_{k=1}^K\sum_{j:(j,k)\in\mathcal{A}}||\mathbf{U}^\mathrm{H}_k\mathbf{H}_{kj}\mathbf{V}_j||^2_{\mathrm F}$ is non-increasing in every round of update. Further noting that $\sum_{k=1}^K\sum_{j:(j,k)\in\mathcal{A}}||\mathbf{U}^\mathrm{H}_k\mathbf{H}_{kj}\mathbf{V}_j||^2_{\mathrm F}\ge 0$, Algorithm~\ref{alg:IA} must converge to a local optimum. This completes the proof.


\end{document}